\documentclass[12pt, letterpaper]{JHEP3}

\usepackage{amssymb, amsmath, amsopn, amsthm}

\usepackage{epsfig}

\usepackage{epsfig,multicol}

\usepackage{amsmath}
\usepackage{graphicx}
\usepackage{latexsym}
\usepackage{amsfonts}
\usepackage{amssymb}

\newcommand{\eref}[1]{(\ref{#1})}

\newcommand{\fref}[1]{Figure~\ref{#1}}
\newcommand{\cref}[1]{Chapter~\ref{#1}}
\newcommand{\beq}{\begin{equation}}
\newcommand{\eeq}{\end{equation}}
\newcommand{\ba}{\begin{array}}
\newcommand{\ea}{\end{array}}
\newcommand{\bcenter}{\begin{center}}
\newcommand{\ecenter}{\end{center}}

\def\IB{\relax\hbox{$\inbar\kern-.3em{\rm B}$}}
\def\IC{\relax\hbox{$\inbar\kern-.3em{\rm C}$}}
\def\ID{\relax\hbox{$\inbar\kern-.3em{\rm D}$}}
\def\IE{\relax\hbox{$\inbar\kern-.3em{\rm E}$}}
\def\IF{\relax\hbox{$\inbar\kern-.3em{\rm F}$}}
\def\IG{\relax\hbox{$\inbar\kern-.3em{\rm G}$}}
\def\IGa{\relax\hbox{${\rm I}\kern-.18em\Gamma$}}
\def\IH{\relax{\rm I\kern-.18em H}}
\def\IK{\relax{\rm I\kern-.18em K}}
\def\IL{\relax{\rm I\kern-.18em L}}
\def\IP{\relax{\rm I\kern-.18em P}}
\def\IR{\relax{\rm I\kern-.18em R}}
\def\IZ{\relax\ifmmode\mathchoice
{\hbox{\cmss Z\kern-.4em Z}}{\hbox{\cmss Z\kern-.4em Z}}
{\lower.9pt\hbox{\cmsss Z\kern-.4em Z}}
{\lower1.2pt\hbox{\cmsss Z\kern-.4em Z}}\else{\cmss Z\kern-.4em Z}\fi}
\def\II{\relax{\rm I\kern-.18em I}}

\def\sCC{{\kern 0.27em\vrule height1.45ex width0.03em depth0em
          \kern-0.30em\rm C}}
\def\C{{\mathchoice
  {\sCC}
  {\sCC}
  {\kern 0.225em \vrule height1.05ex width0.025em depth0em \kern-0.25em \rm C}
  {\kern 0.180em \vrule height0.78ex width0.02em depth0em \kern-0.2em \rm C}
        }}
\def\sHH{{\rm I\kern-.16em{}H}}
\def\H{{\mathchoice
  {\sHH}
  {\sHH}
  {\rm I\kern-.13em{}H}
  {\rm I\kern-.13em{}H} }}
\def\sNN{{\rm I\kern-.16em{}N}}
\def\N{{\mathchoice
  {\sNN}
  {\sNN}
  {\rm I\kern-.12em{}N}
  {\rm I\kern-.10em{}N} }}
\def\sPP{{\rm I\kern-.16em{}P}}
\def\P{{\mathchoice
  {\sPP}
  {\sPP}
  {\rm I\kern-.12em{}P}
  {\rm I\kern-.10em{}P} }}
\def\sQQ{{\kern 0.27em \vrule height1.45ex width0.03em depth0em
          \kern-0.30em \rm Q}}
\def\Q{{\mathchoice
        {\sQQ}
        {\sQQ}
  {\kern 0.225em \vrule height1.05ex width0.025em depth0em \kern-0.25em \rm Q}
  {\kern 0.180em \vrule height0.78ex width0.020em depth0em \kern-0.20em \rm Q}
        }}
\def\sRR{{\rm I\kern-0.16em{}R}}
\def\R{{\mathchoice
  {\sRR}
  {\sRR}
  {\rm I\kern-0.12em{}R}
  {\rm I\kern-0.10em{}R} }}
\def\sZZ{{\rm Z\kern-0.32em{}Z}}
\def\Z{{\mathchoice
  {\sZZ}
  {\sZZ} 
  {\rm Z\kern-0.3em{}Z}     
  {\rm Z\kern-0.25em{}Z} }}  
\def\ZZZ{{\rm Z\kern-0.24em{}Z}}
\def\sII{{\rm I\kern-0.16em{}I}}
\def\I{{\mathchoice
  {\sII}
  {\sII}
  {\rm I\kern-0.12em{}I}
  {\rm I\kern-0.10em{}I} }}

\def\inbar{\,\vrule height1.5ex width.4pt depth0pt}
\font\cmss=cmss10 \font\cmsss=cmss10 at 7pt

\def\smiley{\hbox{\large$\bigcirc$\hspace{-0.80em}\raise.2ex
\hbox{$\cdot\cdot$}\kern-.61em\lower.2ex\hbox{\scriptsize$\smile$}}\ }
\def\frowny{\hbox{\large$\bigcirc$\hspace{-0.80em}\raise.2ex
\hbox{$\cdot\cdot$}\kern-.635em\lower.2ex\hbox{\scriptsize$\frown$}}\ }

\def\I{{\rlap{1} \hskip 1.6pt \hbox{1}}}

\makeatletter
\let\hangafter\@hangfrom
\makeatother

\def\makeatletter{\catcode`\@=11}
\makeatletter
\def\mathbox#1{\hbox{$\m@th#1$}}%
\def\math@ccstyles#1#2#3#4#5#6#7{{\leavevmode
     \setbox0\mathbox{#6#7}%
     \setbox2\mathbox{#4#5}%
     \dimen@ #3%
     \baselineskip\z@\lineskiplimit#1\lineskip\z@
     \vbox{\ialign{##\crcr
            \hfil \kern #2\box2 \hfil\crcr
            \noalign{\kern\dimen@}%
            \hfil\box0\hfil\crcr}}}}
\def\mathaccstyles{\math@ccstyles\maxdimen}
\def\maththroughstyles{\math@ccstyles{-\maxdimen}}
\def\unity%
{\maththroughstyles{.45\ht0}\z@\displaystyle {\mathchar"006C}\displaystyle 1}

\newcommand{\nn}{\nonumber}

\renewcommand{\H}{\mathcal{H}}

\newcommand{\be}{\begin{eqnarray}}
\newcommand{\bea}{\begin{eqnarray}}
\newcommand{\ee}{\end{eqnarray}}
\newcommand{\eea}{\end{eqnarray}}

\newcommand{\bb}{\mathsf{b}}
\newcommand{\ww}{\mathsf{w}}

\title{Towards the Continuous Limit of Cluster Integrable Systems}

\author{Sebasti\'an Franco$^{1,2}$, Daniele Galloni$^2$ and Yang-Hui He$^{3,4,5}$

\\

\vspace{0.35cm}
~\\
$^1$ Theory Group, SLAC National Accelerator Laboratory \\
Menlo Park, CA 94309, USA \\
\vspace{0.25cm}

$^2$ Institute for Particle Physics Phenomenology, Department of Physics \\
Durham University, Durham DH1 3LE, United Kingdom \\
\vspace{0.25cm}

$^3$ Department of Mathematics, City University, London, EC1V 0HB \\
United Kingdom \\
\vspace{0.25cm}

$^4$ School of Physics, NanKai University, Tianjin, 300071, P.R. China \\
\vspace{0.25cm}

$^5$ Merton College, University of Oxford, OX14JD, United Kingdom \\

\vspace{0.25cm}

\email{sfranco@slac.stanford.edu, daniele.galloni@durham.ac.uk, Yang-Hui.He.1@city.ac.uk}\\

}

\abstract{We initiate the study of how to extend the correspondence between dimer models and (0+1)-dimensional cluster integrable systems to (1+1) and (2+1)-dimensional continuous integrable field theories, addressing various points that are necessary for achieving this goal. We first study how to glue and split two integrable systems, from the perspectives of the spectral curve, the resolution of the associated toric Calabi-Yau 3-folds and Higgsing in quiver theories on  D3-brane probes. We identify a continuous parameter controlling the decoupling between the components and present two complementary methods for determining the dependence on this parameter of the dynamical variables of the integrable system. Interested in constructing systems with an infinite number of degrees of freedom, we study the combinatorics of integrable systems built up from a large number of elementary components, and introduce a toy model capturing important features expected to be present in a continuous reformulation of cluster integrable systems.}

\preprint{IPPP/12/15, DCPT/12/30}

\begin{document}

\tableofcontents

\section{Introduction}

It has been more than a decade since the emergence of the technology of singularity resolution by D-branes, in the construction of supersymmetric gauge theories on their worldvolume.
This connection between gauge theory dynamics and Calabi-Yau geometry has been a triumph of the AdS/CFT correspondence \cite{Maldacena:1997re,Gubser:1998bc,Witten:1998qj}.
Much progress has undoubtedly been made, especially for the largest class known to the correspondence, constituting almost all the known examples.
This is the class of toric singularities \cite{Douglas:1997de,Beasley:1999uz,Feng:2000mi,Feng:2001bn}.

For four-dimensional gauge theories arising from toric Calabi-Yau threefolds, the geometry is controlled by a planar toric diagram consisting of a convex, lattice polygon and its interior lattice points.
On the other hand, it has by now become clear that the best way to understand the physics is through dimer models or, equivalently, periodic brane tilings on the plane \cite{Hanany:2005ve,Franco:2005rj,Franco:2005sm,Feng:2005gw}.
The complete encoding of the worldvolume quiver gauge theory in terms of these bi-partite graphs drawn on a torus has led to both the simplification of existing problems as well as development of new ideas in a plethora of directions ranging from physics to mathematics.
The computation of moduli spaces has been trivialized, quantities such as R-charges or procedures such as Higgsing or Seiberg duality \cite{Seiberg:1994pq} have found new graphical realizations. In parallel, geometric resolution by combinatorics of perfect matchings or number-theoretic issues such as the field extension of the defining torus have also provided a fruitful dialogue.

Last year, the path to a new direction was illuminated in the mathematics literature by \cite{GK}, where a correspondence between dimer models and certain integrable models was found.
The correspondence associates any dimer model on a torus to a $(0+1)$-dimensional quantum integrable system -- which they dub {\it cluster integrable system} -- whose phase space contains the moduli space of line bundles with connections on $\Gamma$ and whose Hamiltonian and Casimir operators are given by the partition function of the dimer model.

This connection was readily exploited in our current context of toric gauge theories in \cite{Franco:2011sz,Eager:2011dp,Amariti:2012dd}.
Explicit integrable systems can be thus established for the myriad of theories in the catalogue (cf.~\cite{Davey:2009bp,Hanany:2012hi}) of brane tilings.

Fortified by the strengthening of this correspondence in various guises, we propose to extend it in yet another direction, investigating whether a certain continuous limit exists. Indeed, our dimer models are finite graphs, thereby giving a finite number of faces in the fundamental domain of the torus. Loops around these faces, together with loops around the two fundamental directions of the torus, provide a parametrization of the Poisson manifold of the integrable system. Nevertheless, one could envision a toric diagram with an infinite number of lattice points corresponding to some singularity of infinite order -- obtained for example by taking the limit of an Abelian orbifold with the quotient group taken to infinite order. 
In this situation, we shall be dealing with an integrable system with an infinite number of degrees of freedom. Can we construct in this way continuous integrable field theories in (1+1) or even (2+1) dimensions? If so, can dimer models provide new tools or useful perspectives?

In order to initiate this study we need to understand how to ``glue'' and ``split'' two integrable systems in our present framework. The procedure of splitting or gluing toric diagrams, in relation to geometric resolutions, is well-known. An efficient method for studying general resolutions exploiting dimer models was introduced \cite{GarciaEtxebarria:2006aq}.
In this paper we will first systematically study how this splitting/gluing is mapped to the integrable-system variables, and will demonstrate how this manifests in the spectral curve (defined as the zero locus of the Newton polynomial) associated to the toric diagram. We will also discuss the combinatorics of contributions to conserved charges in the limit in which a large number of building blocks are pasted. Finally, we will introduce a toy model that exhibits some of the desired features of a reformulation of cluster integrable systems in the continuous limit.

The organization of the paper is as follows. In Section \ref{section_dimer_cluster_IS} we review the correspondence between dimer models and integrable systems. Section \ref{section_gluing_and_splitting} discusses the gluing and splitting of spectral surfaces and its relation to the resolution of toric singularities. In \ref{section_continuous_parameter} we identify a continuous parameter controlling the separation of individual components. We relate this parameter to expectation values in the underlying quiver gauge theory and introduce two methods for finding the dependence of the integrable system on it. These ideas are illustrated with explicit examples in Section \ref{section_examples}. In Section \ref{section_large-N}, we investigate the combination of a large number of building blocks and introduce a simple toy model that exhibits some properties we expect in the continuous limit of cluster integrable systems. We conclude in Section \ref{section_conclusions}.

\bigskip

\section{Dimer Models and Cluster Integrable Systems}

\label{section_dimer_cluster_IS}

A remarkable correspondence linking dimer models to an infinite class of integrable systems, denoted {\it cluster integrable systems}, was recently introduced in \cite{GK}. We now provide a brief review of the correspondence.

The Poisson manifold of the integrable system is parametrized by oriented loops on the brane tiling. Cycles going clockwise around each face $w_i$ ($i=1,\ldots,N_g$, with $N_g$ the number of gauge groups in the quiver) and the cycles $z_1$ and $z_2$ wrapping the two directions of the 2-torus provide one possible basis for loops. The $w$-variables are subject to the constraint $\prod_{i=1}^{N_g} w_i=1$, which sometimes can be used to simplify expressions.
 
The Poisson brackets are given by
\beq
\begin{array}{ccl}
\{w_i,w_j\} & = & \epsilon_{w_i,w_j} \, w_i w_j \\
\{z_1,z_2\} & = & 1+\epsilon_{z_1,z_2} \\
\{z_a,w_i\} & = & \epsilon_{z_a,w_i} 
\end{array}
\label{Poisson_brackets}
\eeq
where $\epsilon_{x,y}$ is the number of edges on which the $x$ and $y$ loops overlap, with orientation. Then, $\epsilon_{w_i,w_j}$ is simply the antisymmetric oriented adjacency matrix that counts the number of arrows between gauge groups in the quiver dual to the brane tiling.

The integrable system can be quantized replacing the Poisson brackets by a q-deformed algebra, which takes the form
\beq
X_i X_j = q^{n_{ij}} X_j X_i \, ,
\label{commutators}
\eeq
where $X_i=e^{x_i}$, $q=e^{-i2\pi \hbar}$ and $n_{ij}=\{x_i,x_j\}/(x_i x_j)$.

Every perfect matching is associated with a point in a toric diagram \cite{Franco:2005rj}. When more than one perfect matching corresponds to the same point, we must add all their contributions. As we have just reviewed, the natural variables of cluster integrable systems are loops. We can translate any perfect matching into a closed loop on the tiling by subtracting a reference perfect matching.

In \cite{GK}, it was shown that the commutators defined by \eref{commutators} and \eref{Poisson_brackets} give rise to $(0+1)$-dimensional quantum integrable system, whose conserved charges are:

\medskip

\begin{itemize}
\item{\bf Casimirs:} they commute with everything and are given by the ratio between contributions associated to consecutive points on the boundary of the toric diagram.

\item{\bf Hamiltonians:} they commute with each other and correspond to the internal points in the toric diagram. 
\end{itemize}

\medskip

The toric diagram of the Calabi-Yau 3-fold associated to the dimer model gives rise to a Riemann surface of genus equaling to the number of internal points.
The latter is given by the zero locus of the Newton or characteristic polynomial
\begin{equation}\label{spec}
D = \{ (v_1, v_2)_i \} \quad \leadsto \quad
P(z_1,z_2) = \sum\limits_i c_i z_1^{v_1} z_2^{v_2} \ ,
\end{equation}
where $D$ is the planar toric diagram, specified by lattice points $(v_1, v_2)_i$ and $z_1, z_2 \in 
\IC$ are complex coordinates. The coefficients $c_i$ are functions of the $w$-variables. This Riemann surface is indeed the {\it spectral curve} of the integrable system.

The full Poisson manifold of the integrable system is obtained by gluing different patches via cluster transformations, equivalently Seiberg duality in the associated quiver gauge theories.

\bigskip

\section{Gluing and Splitting}

\label{section_gluing_and_splitting}

In this section we discuss the decomposition of a spectral curve into pieces and the reverse procedure of gluing spectral curves. We also consider another context in which this process arises, the desingularization, or Higgsing, of a Calabi-Yau space. The intimate connection between the decomposition of integrable systems and Higgsing will be the topic of forthcoming sections.

\subsection{Spectral Curves}

\label{section_splitting_spectral_curves}

\begin{figure}[h]
 \centering
 \begin{tabular}[c]{ccc}
 \epsfig{file=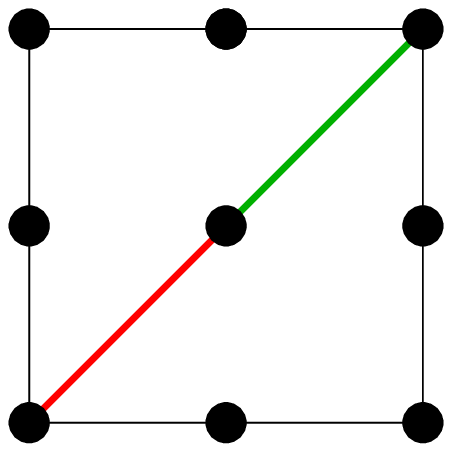,width=0.25\linewidth,clip=} & \ \ \ \ &
\epsfig{file=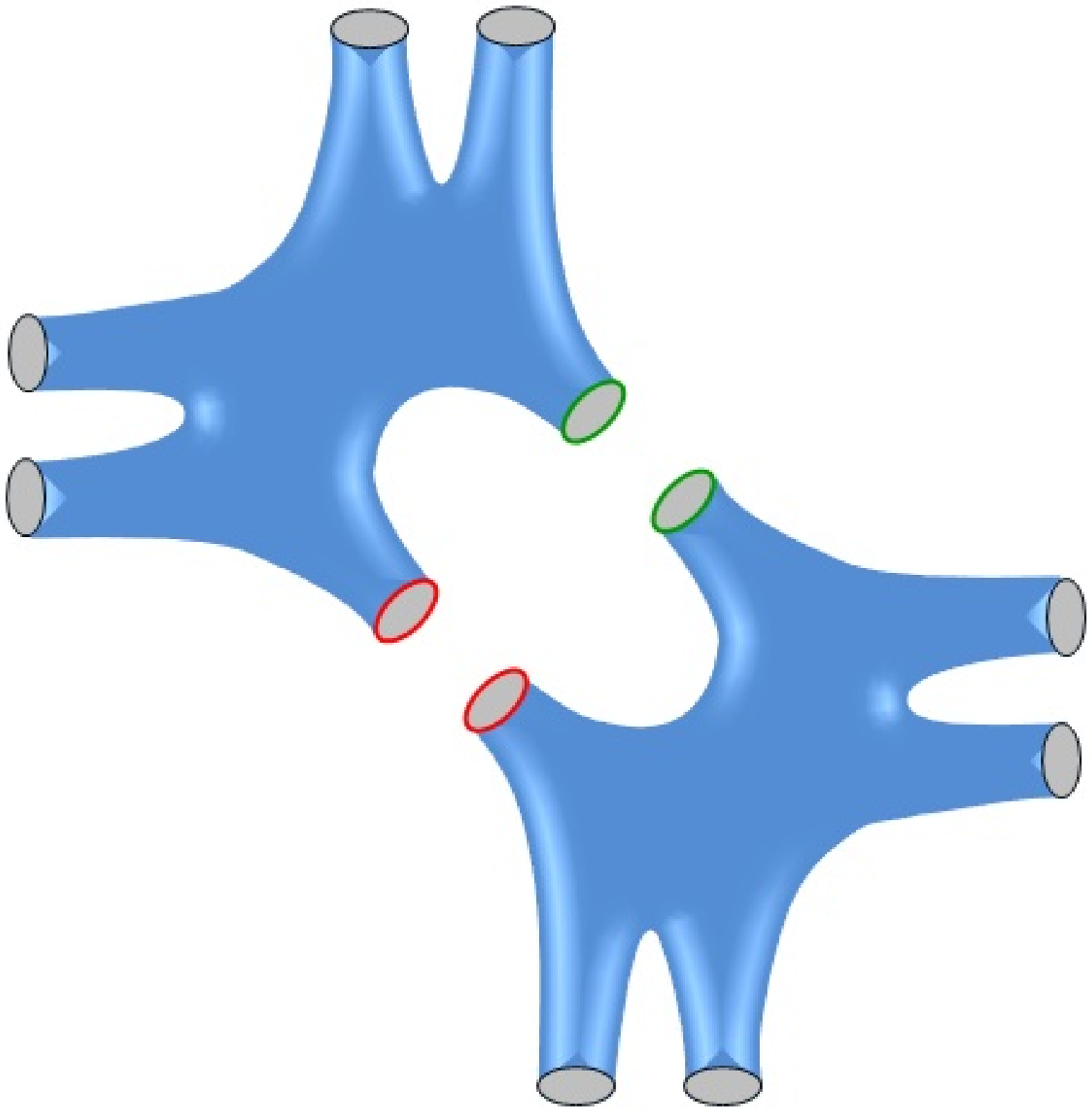,width=0.4\linewidth,clip=} 
 \end{tabular}
\caption{Splitting of a Riemann surface into two daughters.}
\label{2_sigmas} 
\end{figure}

Let us consider the splitting process
\begin{equation}\label{spec-split}
\Sigma \longrightarrow \Sigma_1 + \Sigma_2 \ ,
\end{equation}
where $\Sigma$ is the ``parent'' spectral curve and $\Sigma_{1,2}$ are  the two daughters. In this process we elongate certain throats of the $\Sigma$ until it breaks into two pieces. Since the spectral curve is a thickening of the $(p,q)$-web \cite{Aharony:1997ju,Aharony:1997bh}, which is the graph dual to the toric diagram, the connecting throats are dual to segments joining points of the toric diagram along the boundary between the daughters, as shown in \fref{2_sigmas}.

The separation between components of the spectral curve is achieved by tuning the coefficients $c_i$ in the characteristic polynomial \eref{spec}. In the integrable system, these coefficients are functions of the $w$-variables. In the limit of large distance between components, the $c_i$'s scale differently with respect to the separation and develop a hierarchical structure. 

Given a decomposition of the spectral curve into two pieces, it is natural to expect that in the limit of infinite separation the original integrable system reduces to the sum of the integrable systems associated to the two components. In Sections \ref{section_continuous_parameter} and \ref{section_examples}, we provide a detailed explanation of how this intuition is realized.

\begin{figure}[h]
 \centering
 \begin{tabular}[c]{ccc}
 \epsfig{file=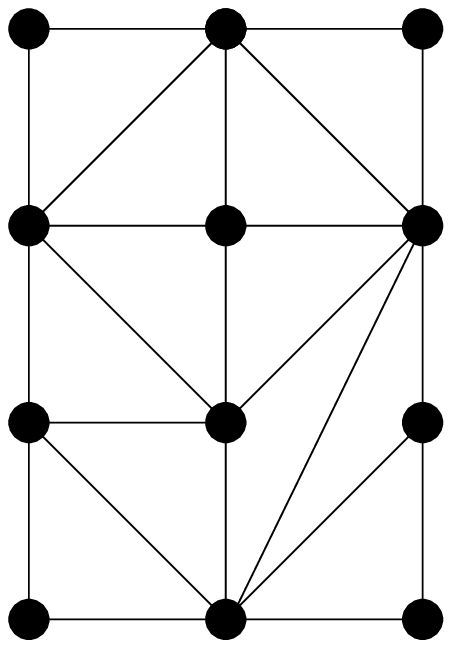,width=0.15\linewidth,clip=} & \ \ \ \ &
\epsfig{file=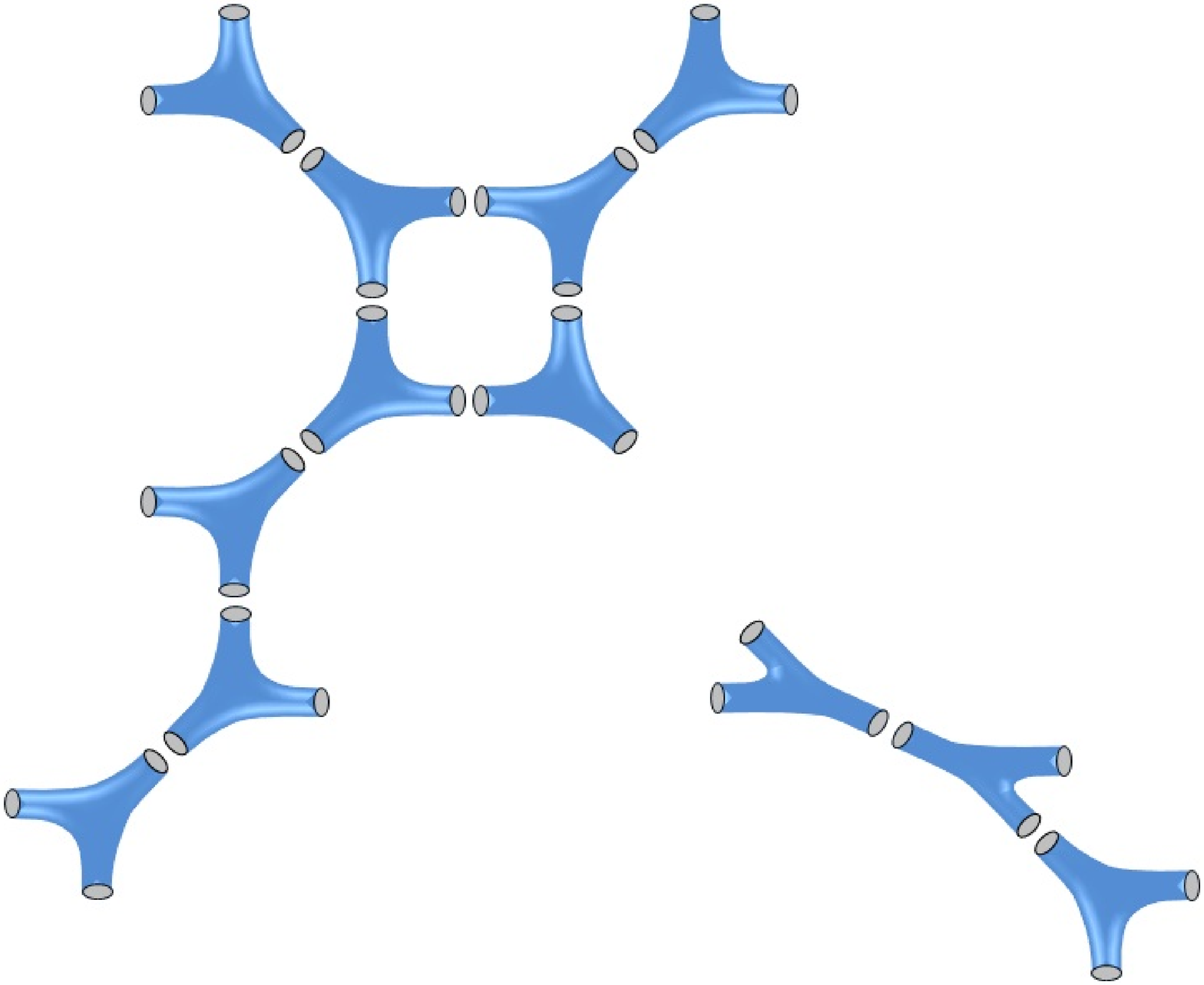,width=0.45\linewidth,clip=} 
 \end{tabular}
\caption{A general triangulation of the toric diagram and the corresponding trinion decomposition of the spectral curve.}
\label{trinion_decomposition} 
\end{figure}

Taking the splitting process to an extreme we obtain a decomposition of any Riemann surface into a collection of {\it trinions}, i.e. spheres with three punctures. Any such decomposition is in one-to-one correspondence with triangulations of the toric diagram. Each triangle gives rise to a trinion. Their number is thus equal to twice the area of the toric diagram, which in turn is equal to the number of gauge groups in the associated quiver gauge theory. These decompositions allow us to see the full integrable system continuously emerge from the combination of trivial integrable systems associated to trinions. \fref{trinion_decomposition} shows a possible triangulation of a toric diagram and its corresponding trinion decomposition.

A standard method in real (``tropical'') geometry to visualize Riemann surfaces, which will prove useful in later sections, is the so-called {\bf am{\oe}ba projection}:
\begin{equation}
\mathcal{A} \ : \ (z_1, z_2) \mapsto (\log |z_1|, \log |z_2|) \, .
\end{equation}
The am{\oe}ba can be thought of as a thickening of the graph-dual of the toric diagram; i.e., we can draw the $(p,q)$-web from $D$ and this will constitute the ``spine'' of the am{\oe}ba. In the actual plot, the ``tentacles'' which tend to infinity will have their directions given by the $(p,q)$-vectors which are normal to the toric diagram.

\begin{figure}[h]
\begin{center}
\includegraphics[width=12cm]{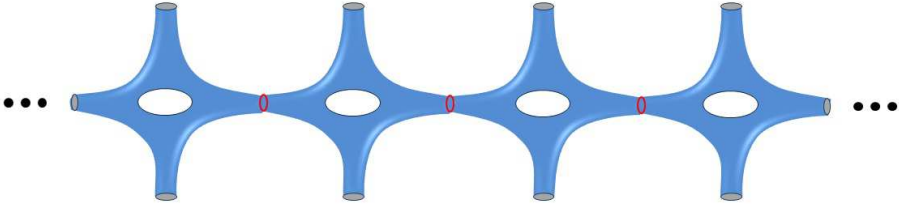}
\caption{Combining an infinite number of elementary spectral curves to generate a $(1+1)$-dimensional theory. By performing a similar gluing along the vertical direction we expect to generate a $(2+1)$-dimensional theory.}
\label{continuous_limit}
\end{center}
\end{figure}

As explained in the introduction, what we ultimately wish to study is the integrable system that emerges in the continuous limit where we glue a countably infinite number of toric sub-diagrams, or equivalently, spectral curves. We conjecture that, depending on how we assemble these building blocks, the integrable systems with an infinite number of degrees of freedom that are generated by this procedure are $(1+1)$ or $(2+1)$-dimensional integrable field theories. \fref{continuous_limit} shows a number of elementary spectral curves glued to generate a $(1+1)$-dimensional theory.

\subsection{Partial Resolution of Calabi-Yau Singularities and Higgsing}

The splitting and gluing of Riemann surfaces of the type discussed in Section \ref{section_splitting_spectral_curves} also play an important role in the context of partial resolutions of toric singular Calabi-Yau 3-folds. We now review this process from geometric, gauge theoretic and dimer model perspectives. In Section \ref{section_continuous_parameter} we elaborate on the intimate connections with the (de)composition of integrable systems.

Geometrically, the partial resolution of a toric singularity corresponds to the process illustrated in \fref{2_sigmas}: one takes the toric diagram and divides it into components \footnote{In the dual cone picture of the toric variety, this is the process of stellar division \cite{fulton}.}. 
The resulting components must correspond to well-behaved toric diagrams, which constrain them to be convex. 
For concreteness, we will focus on the case in which we split the toric diagram into two parts, to which we shall affectionately refer as the parent with two daughters. It is possible to deal with more components by iteration of this procedure.\footnote{Not all decompositions can be reduced to a sequence of binary splittings. A necessary condition is that, at each step, the toric diagrams of the daughters are convex.}

Let us now describe the resolution from the perspective of the quiver theory on the worldvolume of D3-brane probes. This is of course standard technology dating back to the early days of studying quiver gauge theories from toric Calabi-Yau singularities \cite{Douglas:1997de,Beasley:1999uz,Feng:2000mi,Feng:2001bn}.
The starting point is a set of $N=n_1+n_2$ D3-branes on the parent singularity. All chiral fields in the parent quiver are  $(n_1+n_2)\times (n_1+n_2)$ matrices. 
The parent singularity is then resolved into two daughter singularities containing $n_1$ and $n_2$ D3-branes, respectively.
As a result, we obtain two decoupled quiver gauge theories, whose gauge group ranks are given by $n_1$ and $n_2$.
From a gauge theory viewpoint, this resolution corresponds to turning on non-zero vacuum expectation values (vevs) for some block sub-matrices in the scalar components of these fields. Fields charged under gauge groups in both quivers have masses controlled by the expectation values and decouple from the low energy theory. 

\begin{figure}[h]
\begin{center}
\includegraphics[width=7.5cm]{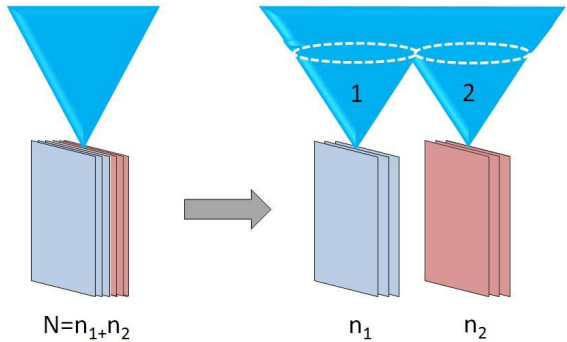}
\caption{The resolution of the parent singularity with parallel coincident $N=n_1+n_2$ D3-branes (on the left) results in two daughter theories with $n_1$ and $n_2$ D3-brane respectively.
The blue cone signifies the Calabi-Yau singularity at the tip of which the D3-branes sit.}
\label{resolution_branes}
\end{center}
\end{figure}

A pictorial representation of this process is given in \fref{resolution_branes}. One could envisage, of course, the reverse process of gluing to produce a more singular parent. This should correspond to an un-Higgsing mechanism (q.~v.~\cite{Feng:2002fv}). We will illustrate these ideas with explicit examples in Section \ref{section_examples}.

For our purposes, it is sufficient to focus on the simple case in which $n_1=n_2=1$. We focus on diagonal vevs of the form
\beq
\langle X_{ij}\rangle = \left(\begin{array}{cc} X_{ij}^{(1)} & 0 \\ 0 & X_{ij}^{(2)}\end{array}\right) \, ,
\label{matrix_vevs}
\eeq
for the field $X_{ij}$ in the parent theory. We will stick to this case throughout the paper. Resolutions generically involve turning on several non-zero expectation values simultaneously. We restrict to the case in which all non-zero vevs have the same magnitude, which will turn out to control the distance between the daughter singularities.\footnote{Theories with different vevs give rise to multiple energy scales. If these scales are hierarchically separated, the Higgsing process can be studied sequentially.} The acquisition of vevs in this fashion will split the parent theory into its two daughters.

\bigskip

\subsubsection{Resolution in the Dimer Model}

\label{section_resolution_dimer}

One of the greatest computational challenges to the resolution of toric singularities by D3-branes was the identification of which fields in the parent acquire non-zero vevs \cite{Feng:2000mi,Feng:2001bn}.
This issue was resolved by the dimer model representation of toric quiver gauge theories \cite{Hanany:2005ve,Franco:2005rj,Feng:2005gw}.
Dimer models are extremely useful for identifying the non-zero vevs that are necessary in order to achieve a given resolution. An elegant description of general partial resolutions, exploiting the map between the dimer model and a tiling of the spectral curve, was introduced in \cite{GarciaEtxebarria:2006aq}. We now briefly review this procedure.

Like perfect matchings, zig-zag paths play a prominent role in connecting dimer models to geometry. They are defined as paths that alternate between turning maximally right and maximally left at consecutive nodes in the bane tiling. Each edge, then, has exactly two oppositely oriented zig-zag paths, criss-crossing before heading to nodes of opposite color, weaving an 
intertwined pattern on the torus. For consistent gauge theories, these zig-zag paths never intersect themselves and form closed loops wrapping $(p,q)$-cycles on the torus.

The {\it untwisting map} is an operation on zig-zag paths that exchanges the criss-cross and turns a brane tiling, which by construction lives on a 2-torus, into a tiling of the spectral curve $\Sigma$ (which we recall is the zero locus of the Newton polynomial for the given toric diagram) and vice versa. We refer the reader to \cite{Feng:2005gw} for a detailed explanation of the untwisting map whose effect on zig-zag paths of both $\mathbb{T}^2$ and $\Sigma$ is summarized below.
\bigskip
\begin{center}
\begin{tabular}{ccc}
$\mathbb{T}^2$ & & $\Sigma$ \\ \hline
zig-zag path & $\leftrightarrow$ & \ \ \ face = puncture \ \ \ \\
\ \ \ face = gauge group \ \ \ & $\leftrightarrow$ & zig-zag path
\end{tabular}
\end{center}
\bigskip

Starting from the parent spectral curve $\Sigma$, we elongate one or several internal throats that connect the two daughters, $\Sigma_1$ and $\Sigma_2$. The daughters then decouple in the limit in which these throats become infinitely long. From the viewpoint of the daughters, these throats become new external legs, i.e., new punctures.
The appearance of new punctures can easily be implemented in terms of brane tilings. We consider one copy of the original brane tiling on $\mathbb{T}^2$ for each of the two daughters. On each copy, we draw the zig-zag paths associated to the original punctures that will end up on the corresponding component. Next, we introduce the paths which are the complement to these zig-zag paths in the original set. These new paths correspond to the new punctures that are generated in the splitting process. 

In order for the new paths to become actual zig-zag paths, some edges must be removed from the daughter tilings. The bifundamentals on the tiling that do not have any paths running over them are removed. These are precisely the ones that acquire non-zero vevs in the Higgsing. Generically, different edges are removed from the two tilings associated to the daughters. This is the manifestation, in dimer language, of the matrix vevs in \eref{matrix_vevs}. In Section \ref{section_examples}, we present explicit examples illustrating this procedure. Bivalent nodes might be generated when removing edges. They correspond to massive fields that can be integrated out \cite{Franco:2005rj}. From the perspective of the daughter integrable systems, the generation of new punctures corresponds to the appearance of new Casimir operators.

When Higgsing, the rule for removing perfect matchings is simple: every perfect matching containing an edge corresponding to a field with a non-zero vev, must be eliminated. The mapping of the dimer model into a tiling of the spectral curve provides an interesting alternative perspective on the disappearance of perfect matchings. Differences of perfect matchings are translated into 1-cycles with appropriate homology on $\Sigma$. The difference between two adjacent perfect matchings, say $p$ and $p'$, in the toric diagram simply corresponds to a 1-cycle wrapped around the associated elongation of $\Sigma$. Once $\Sigma$ has been split, certain 1-cycles will no longer be able to exist.  
If the difference between two perfect matchings is contained in any of the sub-dimers, both perfect matchings will survive in the corresponding component of the daughter singularity. On the other hand, if the difference is not contained in any sub-dimer, one or both of the perfect matchings will not survive the Higgsing process. 

In general, the set of non-zero expectation values resulting in a given decomposition, or, equivalently, the set of removed edges from the daughter brane tilings, is not unique. In Section \ref{section_examples}, we will discuss the issue of multiple solutions for an explicit example in detail.

\medskip

\section{A Continuous Control Parameter}

\label{section_continuous_parameter}

As we have already mentioned, the splitting of the spectral curve follows from certain hierarchies between the coefficients in the characteristic polynomial. These hierarchies are controlled by a continuous parameter, which we will denote $\Lambda$; in quiver language it is connected to the non-zero expectation values of bifundamental fields.

In this section, we introduce two complementary approaches for determining the precise dependence of $\Lambda$ on the coefficients of $P(z_1,z_2)$ that achieves a given $\Sigma \to \Sigma_1+\Sigma_2$ decomposition. 

\bigskip

\subsection{Scalings from VEVs}

Without loss of generality, we can identify $\Lambda$ with the non-zero vevs, i.e. $\langle X_{ij}^{(a)}\rangle=\Lambda$. The coefficients in $P(z_1,z_2)$ are polynomials in the $w_i$ variables, corresponding to closed loops with vanishing homology on the brane tiling. Following \cite{Kenyon:2003uj} (see also \cite{Franco:2006gc} for applications of this idea), we can write any loop in terms of edges in the brane tiling as

\beq
v(\gamma)=\prod_{i=1}^{k-1} {X(\ww_i,\bb_i)\over X(\ww_{i+1},\bb_i)} \, ,
\label{flux_v}
\eeq
where the product runs over the contour $\gamma$ and $\bb_i$ and $\ww_j$ denote black and white nodes. Here, $X(\ww_i,\bb_i)$ and $X(\ww_{i+1},\bb_i)$ are bifundamental fields, in which we explicitly indicate the tiling nodes connected by the corresponding edge when going around $\gamma$ instead of employing the usual notation with subindices for the gauge groups under which they are charged. Going back and forth between the two notations is straightforward.
Furthermore, we remind the reader that the variables defined Section \ref{section_dimer_cluster_IS} for our integrable system, in terms of the notation in \eqref{flux_v}, are
\begin{equation}
w_j = v(\gamma_{w_j}) \ ; \ z_i = v(\gamma_{z_i}) \qquad
j = 1, \ldots, N_g, \ \ i = 1,2 \ .
\end{equation}

We can thus obtain the $\Lambda$-scaling of any loop by plugging the expectation values into \eref{flux_v}. The conclusion is that we obtain the following factors \footnote{Exchanging black and white nodes, an operation that has no effect on the physics, exchanges $\Lambda$ with $\Lambda^{-1}$. This operation does not affect the scalings of individual contributions to conserved charges though, since it also inverts the orientation of all paths.} 
\beq
\begin{array}{ccccc}
\mbox{ Scaling } \Lambda & \mbox{    for each    } & \langle X(\ww_i,\bb_i)\rangle & = & \Lambda \\
\mbox{ Scaling } \Lambda^{-1} & \mbox{    for each    } & \langle X(\ww_{i+1},\bb_i)\rangle & = & \Lambda 
\end{array}
\label{Lambda_weight_loop}
\eeq

\begin{figure}[h!t!]
 \centering
 \begin{tabular}[c]{ccccc}
 \epsfig{file=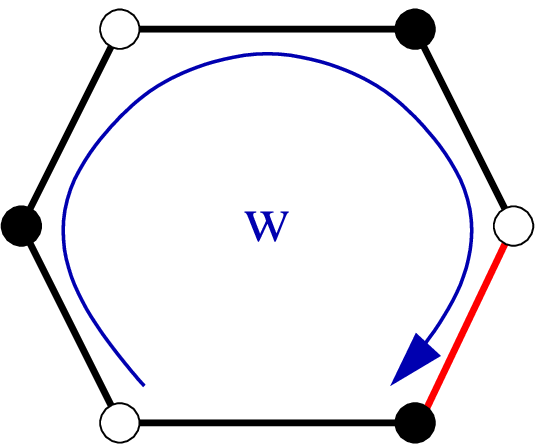,width=0.15\linewidth,clip=} & \ \ \ \ &
\epsfig{file=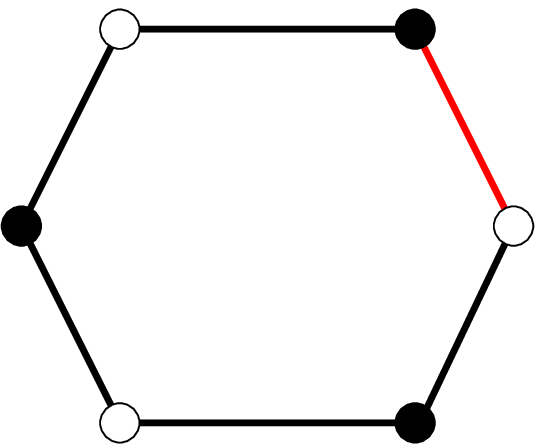,width=0.15\linewidth,clip=} 
& \ \ \ \ &
\epsfig{file=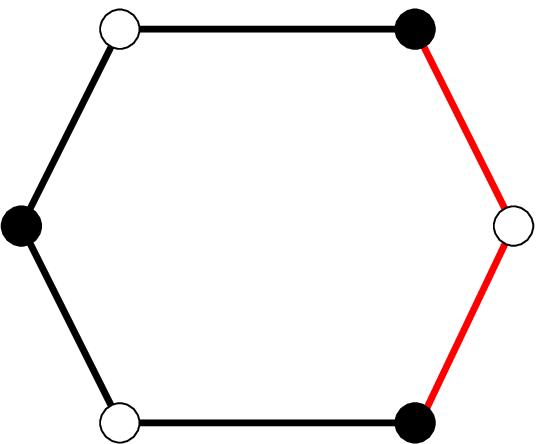,width=0.15\linewidth,clip=} \\
$\Lambda$ & & $\Lambda^{-1}$ & & $\Lambda^0$
 \end{tabular}
\caption{Examples of $\Lambda$ scalings of a $w_i$ cycle, which we have drawn as the cyclic arrow in blue.
We indicate the edge associated with a field with a non-zero vev in red.}
\label{dimer_w_weight} 
\end{figure}

Applying \eref{Lambda_weight_loop} directly to the $w_i$ cycles, we can rephrase it in terms of vevs for fields transforming in the fundamental or antifundamental representation of the corresponding gauge group. We obtain
\beq
\begin{array}{ccccccc}
\underline{\mbox{within } w_i \mbox{ cycle}}: & \ \ & \Lambda & \mbox{    for each    } & \langle X_{ji}\rangle & = & \Lambda \\
& & \Lambda^{-1} & \mbox{    for each    } & \langle X_{ij}\rangle & = & \Lambda 
\end{array}
\label{Lambda_weight_w}
\eeq
where we have changed the notation and subindices indicate gauge groups connected by bifundamentals. We show some examples in \fref{dimer_w_weight}. 

Notice that we have not included color indices in \eref{flux_v}. Similarly, we have not specified on which of the two daughter components, which have different sets of non-zero vevs, the scalings in \eref{Lambda_weight_loop} and \eref{Lambda_weight_w} must be calculated. However, on physical grounds, it is clear that determining the scaling in any of the two daughters leads to the same result. We shall present explicit examples in Section \ref{section_examples}.

We have explained how to determine the $\Lambda$ of any loop in the tiling arising from a general decomposition of the Riemann surface. It is important to emphasize that the $\Lambda$-independent part of these loops can be completely arbitrary in size, but in a way that is uncorrelated with the splitting of $\Sigma$.

\bigskip

\subsection{Scalings from the Spectral Curve}\label{s:algo}

Let us now introduce an alternative method for determining the $\Lambda$ scalings directly associated with a decomposition of the spectral curve in an algorithmic fashion, wherein we obtain a set of explicit Diophantine inequalities which needs to be solved.
Suppose the parent spectral curve, corresponding to the toric diagram $D_P = \{ v_1^i, v_2^i\}$ (with $i$ indexing the nodes) can be written as
\begin{equation}
P(z_1, z_2) = \sum\limits_i \left( \sum_{k=1}^{p_i} \prod\limits_{j=1}^{N_G} w_j^{\alpha_{k,j}^i} \right)  z_1^{v_1^i} z_2^{v_2^i} \ ;
\end{equation}
here, we have explicitly written the coefficients as sums over monomials in the $w$-variables.
Indeed, the number of terms, which we index by $k$, of monomials for the $i$-th node is the number of perfect matchings $p_i$ for that node. Finally, each monomial is a product of $w$-variables, indexed by $j$, raised to integers $\alpha_{k,j}^i$.\footnote{It is always possible if desired, although not necessary, to bring the monomials to a form in which $\alpha_{k,j}^i\geq 0$ by using $\prod_{j=1}^{N_g} w_j=1$.}

Now, when we Higgs, the coefficients -- as polynomials in $w$-variables -- separate into the sum of two types of terms: those which survive the Higgsing, which we shall separate into the first $s_i$ terms for the $i$-th node, and those which do not, which are the remaining $p_i - s_i$ terms:
\begin{equation}\label{split}
P(z_1, z_2) = \sum\limits_i \left( 
\sum_{k=1}^{s_i} 
\prod\limits_{j=1}^{N_G} w_j^{\alpha_{k,j}^i} 
+
\sum_{k=s_i+1}^{p_i} 
\prod\limits_{j=1}^{N_G} w_j^{\beta_{k,j}^i} \right)  
z_1^{v_1^i} z_2^{v_2^i} \ ;
\end{equation}
as an emphasis, we have relabeled the powers of the monomials which do not survive as $\beta_{k,j}^i$.

Now, let us introduce the scalings (we can take, without loss of generality, the powers $k_j$ to be integers):
\begin{equation}
w_j \longrightarrow \Lambda^{\kappa_j} \  , \qquad \kappa_j \in \IZ
\end{equation}
and substitute back into \eqref{split} to give
\begin{equation}\label{split2}
P(z_1, z_2) = \sum\limits_i \left( 
\sum_{k=1}^{s_i} 
\Lambda^{\sum\limits_{j=1}^{N_G} \kappa_j \alpha_{k,j}^i}
+
\sum_{k=s_i+1}^{p_i} 
\Lambda^{\sum\limits_{j=1}^{N_G} \kappa_j \beta_{k,j}^i}
\right)
z_1^{v_1^i} z_2^{v_2^i} \ ;
\end{equation}

It is now clear what has to occur: for the terms which survive, they must have the same order in $\Lambda$ and those which do not, must have strictly less order.
In other words, for each $i$, we must have
$\sum_{j=1}^{N_G} \kappa_j \alpha_{1,j}^i = \sum_{j=1}^{N_G} \kappa_j \alpha_{2,j}^i = \ldots = \sum_{j=1}^{N_G} \kappa_j \alpha_{s_i,j}^i\equiv C$.
This $C$ must be strictly greater than each of $\sum_{j=1}^{N_G} \kappa_j \beta_{k,j}^i$ for $k = s_i +1 , \ldots, p_i$.
Finally, we recall that $\prod_{j=1}^{N_G} w_j = 1$, so that $\sum_{j=1}^{N_G} \kappa_j = 0$. The terms that do not survive are precisely those associated to perfect matchings that are removed by the Higgsing. We can use this identification to determine a priori which terms must have a relative $\Lambda$ suppression.

In summary, we have the following set of Diophantine inequalities in $\kappa$:
for each $i = 1, 2, \ldots, n$ where $n$ is the number of nodes in the toric diagram,
\begin{eqnarray}
\nn
C &=& \sum\limits_{j=1}^{N_G} \kappa_j \alpha_{k,j}^i \mbox{ for all } k = 1, \ldots, s_i \\
\nn
C &>& \sum\limits_{j=1}^{N_G} \kappa_j \beta_{k,j}^i \mbox{ for each } k = s_i+1, \ldots, p_i \\
0 &=& \sum\limits_{j=1}^{N_G} \kappa_j \ .
\label{ineq}
\end{eqnarray}

\smallskip

\section{Explicit Examples}

\label{section_examples}

Having abstractly discussed how the splitting should work by obtaining the weights of the variables either from the acquisition of vevs in the dimer or from the coefficients in the spectral curve, we can now illustrate our proposal in detail with some explicit examples.
In this section, we will first study the cone over the double zeroth Hirzebruch surface and then the space $Y^{4,0}$.
These will give ample demonstration of our technique.
\subsection{Double $F_0$}

\label{section_double_F0}

\begin{figure}[!h]
\begin{center}
\includegraphics[width=6cm]{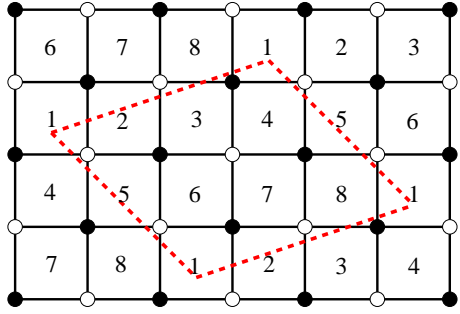}
\caption{Brane tiling for the double $F_0$ theory.}
\label{dimer_2F0}
\end{center}
\end{figure}

\begin{figure}[!h]
\begin{center}
\includegraphics[width=3.3cm]{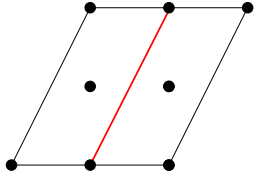}
\caption{Toric diagram for the double $F_0$. A red line indicated how we split it into two components.}
\label{toric_2_F0}
\end{center}
\end{figure}

Let us consider a $\mathbb{Z}_2$ orbifold of $F_0$, whose toric diagram is shown in \fref{toric_2_F0}, to which we refer as the double $F_0$ theory. \fref{dimer_2F0} shows the corresponding brane tiling.
We see that there are 8 the gauge groups, twice that of $F_0$, and 16 bifundamental fields which we shall denote as $X_{ij}$ in standard nomenclature, signifying the field corresponding to the edge bounding face $i$ and face $j$ in \fref{dimer_2F0}.
The superpotential terms are all quartic and can also be instantly read off from the figure. We wish to consider the decomposition of this geometry into two copies of $F_0$, as shown in \fref{toric_2_F0}.

Using the ideas of \cite{GarciaEtxebarria:2006aq}, which were summarized in Section \ref{section_resolution_dimer}, we conclude there are four possible sets of expectation values leading to the same desired decomposition of the geometry. They are:
\beq
\begin{array}{ccccccc}
\mbox{{\bf \underline{Higgsing 1}:}} & \ & X_{41}^{(1)}, X_{12}^{(1)}, X_{85}^{(1)}, X_{56}^{(1)} & \ \ \ \ & \mbox{{\bf \underline{Higgsing 2}:}} & \ & X_{34}^{(1)}, X_{41}^{(1)}, X_{78}^{(1)}, X_{85}^{(1)} \\
& & X_{83}^{(2)}, X_{32}^{(2)}, X_{47}^{(2)}, X_{76}^{(2)} & \ \ \ \ \ & & & X_{32}^{(2)}, X_{25}^{(2)}, X_{76}^{(2)}, X_{61}^{(2)} \\ \\
\mbox{{\bf \underline{Higgsing 3}:}} & & X_{27}^{(1)}, X_{78}^{(1)}, X_{63}^{(1)}, X_{34}^{(1)} & \ \ \ \ & \mbox{{\bf \underline{Higgsing 4}:}} & & X_{12}^{(1)}, X_{27}^{(1)}, X_{56}^{(1)}, X_{63}^{(1)} \\
& & X_{25}^{(2)}, X_{54}^{(2)}, X_{61}^{(2)}, X_{18}^{(2)} & \ \ \ \ & & & X_{18}^{(2)}, X_{83}^{(2)}, X_{54}^{(2)}, X_{47}^{(2)}
\end{array}
\eeq
We have separated, in the above, the 8 fields of the parent theory that get a non-zero vev into its two daughters, which following the notation in \eref{matrix_vevs} we denote by superscripts $(1)$ and $(2)$ respectively. We see that each daughter contains four fields with non-zero vevs.
From these non-zero vevs and the rule prescribed in \eqref{Lambda_weight_w}, we determine the weights of our $w_i$ variables for the four Higgsings:
\beq\label{w-2F0}
\begin{array}{c|cccccccc}
\ \ \mbox{Higgsing} \ \ & \ \ w_1 \ \ & \ \ w_2 \ \ & \ \ w_3 \ \ & \ \ w_4 \ \ & \ \ w_5 \ \ & \ \ w_6 \ \ & \ \ w_7 \ \ & \ \ w_8 \ \ \\ \hline \hline
\mbox{1} & 1 & \Lambda & 1 & \Lambda^{-1} & 1 & \Lambda & 1 & \Lambda^{-1} \\ \hline
\mbox{2} &\Lambda & 1 & \Lambda^{-1} & 1 & \Lambda & 1 & \Lambda^{-1} & 1 \\ \hline
\mbox{3} &1 & \Lambda^{-1} & 1 & \Lambda & 1 & \Lambda^{-1} & 1 & \Lambda \\ \hline
\mbox{4} &\Lambda^{-1} & 1 & \Lambda & 1 & \Lambda^{-1} & 1 & \Lambda & 1 
\end{array}
\eeq

\begin{figure}[h]
 \centering
 \begin{tabular}[c]{ccc}
 \epsfig{file=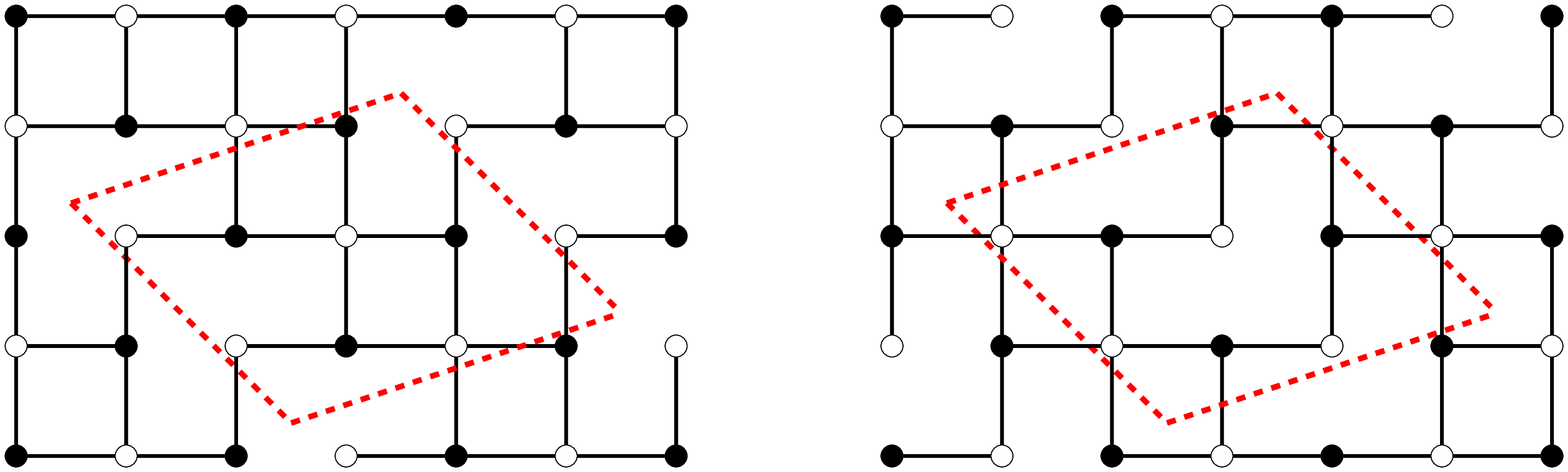,width=0.45\linewidth,clip=} & \ \ \ &
\epsfig{file=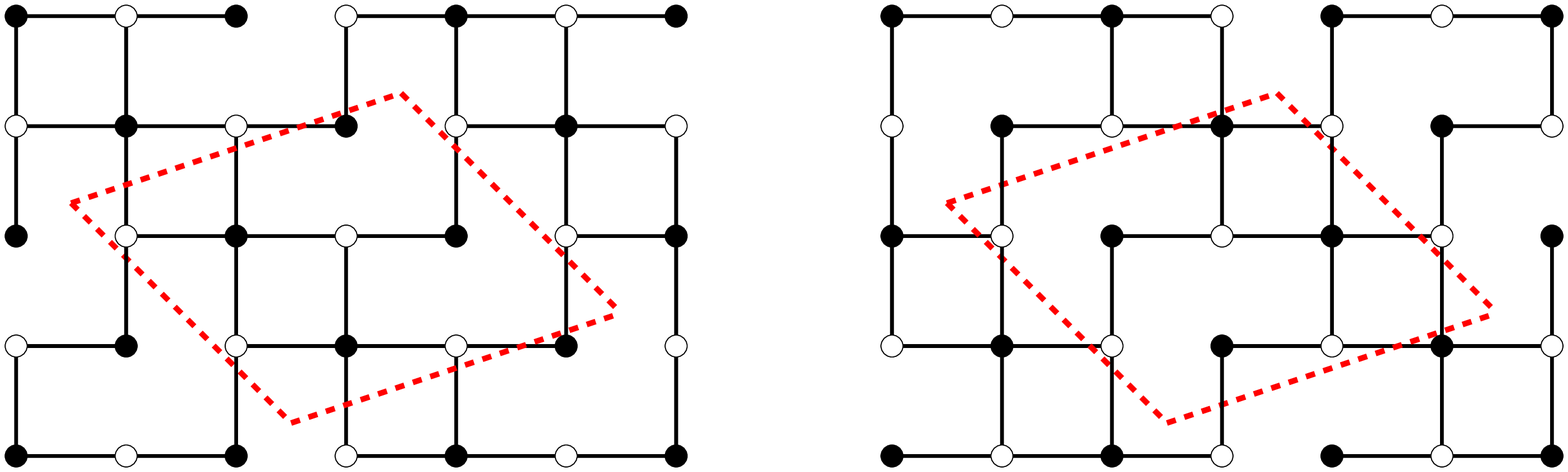,width=0.45\linewidth,clip=} \\ 
\mbox{Higgsing 1} & & \mbox{Higgsing 2} \\ \\
 \epsfig{file=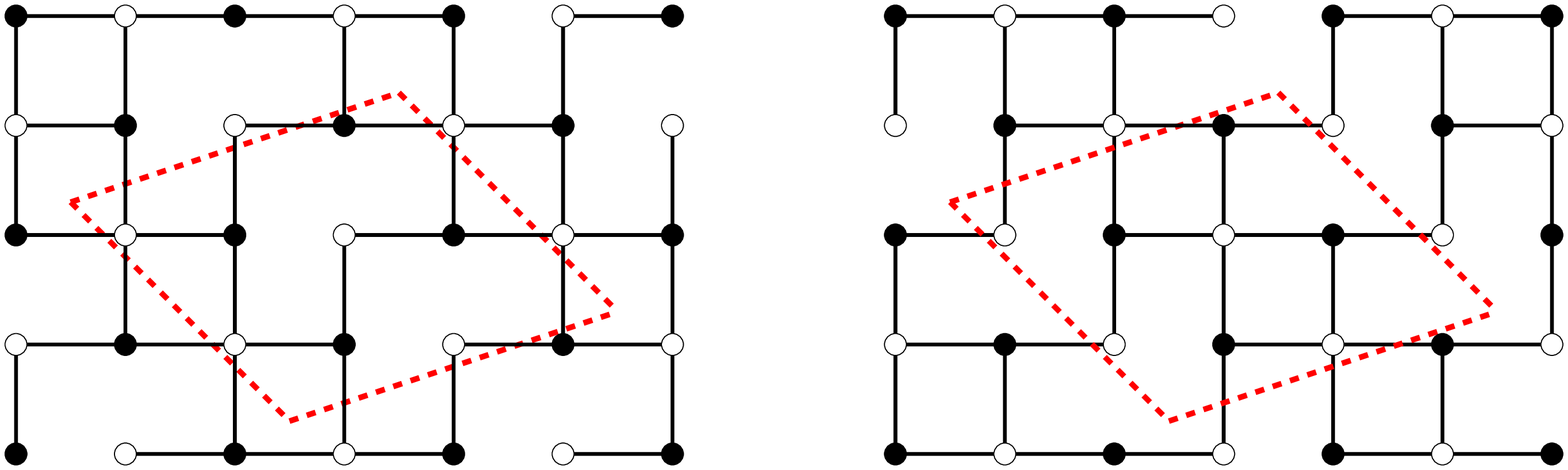,width=0.45\linewidth,clip=} & & 
\epsfig{file=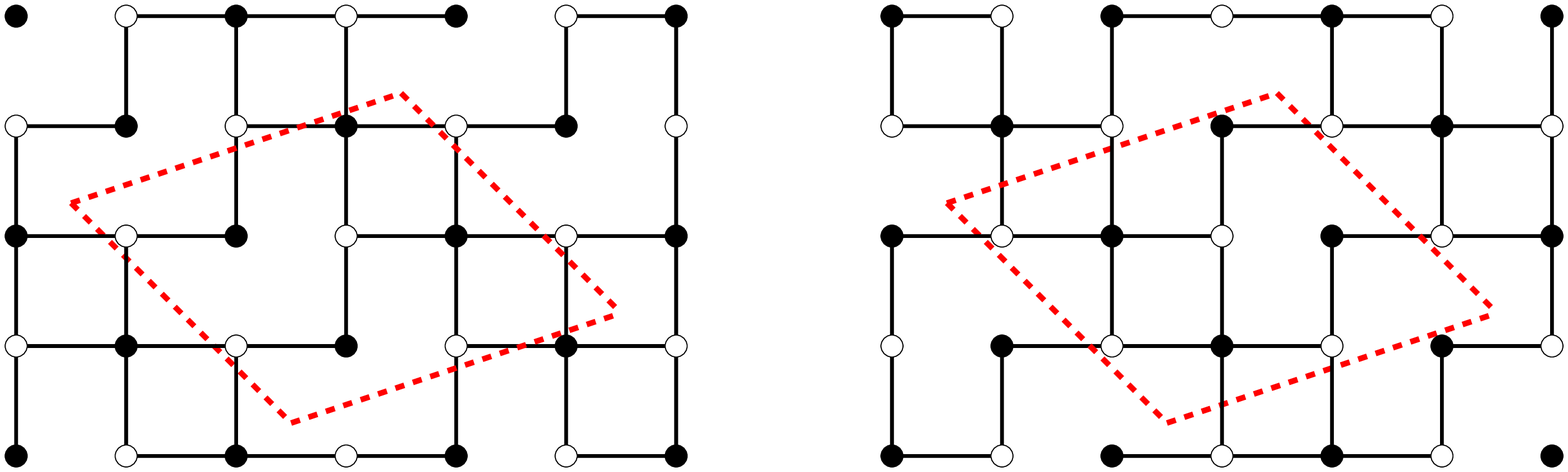,width=0.45\linewidth,clip=} \\
\mbox{Higgsing 3} & & \mbox{Higgsing 4} 
 \end{tabular}
\caption{Dimer models for the four Higgsings of the double $F_0$ theory into two daughter $F_0$ theories.}
\label{dimer_Higgsing_2F0} 
\end{figure}

What is happening in the field theory, as graphically depicted by the dimer, is shown in \fref{dimer_Higgsing_2F0}. In each of the four Higgsings, we separate the parent dimer model into the complementary dimers of the two daughters, (1) on the left and (2) on the right. In order to facilitate comparison with the original parent tiling, we have not integrated out massive fields. If we do so, we obtain the square lattice characteristic of $F_0$.

The weights associated to different Higgsings are simply related by an overall shift, which follows from the fact that the Higgsed tilings are also connected by shifts and rotations as shown in \fref{dimer_Higgsing_2F0}.

\bigskip

\subsubsection*{The Integrable System}

We now investigate the effect of continuously splitting the spectral curve, equivalently Higgsing, on the integrable system. In this process some contributions to conserved charges, those associated perfect matchings removed by the Higgsing, are continuously suppressed until the theory reduces to two decoupled integrable systems.

We can also determine the $\Lambda$-scaling of $w$-variables by analyzing the behavior of coefficients of the spectral curve. Indeed, we can find the weights in \eqref{w-2F0} independently by the procedure outlined in Section  \ref{s:algo}. Reassuringly, the results of using this method agrees in all examples with the ones obtained from \eqref{Lambda_weight_w}. 

The tables below present the information in the integrable system for the double $F_0$ theory. The $(n_1,n_2)$ row refers to the coefficient for the $z_1^{n_1}z_2^{n_2}$ term in the Newton polynomial. For each of the Higgsings, we underline the contributions that survive in the $\Lambda\to \infty$ limit. 
The third column shows the surviving leading $\Lambda$ dependence of the coefficients in the characteristic polynomial.

\bigskip
\bigskip
\bigskip

\noindent{\bf Higgsing 1:}
Let us begin with Higgsing 1 and discuss it in a little more detail.
For each node in the toric diagram, here given as an integer 2-vector, we can specify the coefficient in terms of the $w$-variables within the spectral curve using the technique of \cite{Eager:2011dp}, as reviewed in Section \ref{section_dimer_cluster_IS}.
The number of monomials will correspond to the number of perfect matching for the node.
Let us take the $(1,1)$ point of the toric diagram as an example, which is an internal node with 8 perfect matchings; the term in the spectral curve will be 
\medskip
\[
\left(w_1 w_3 w_4+ w_3 w_4+\underline{w_3}+ \underline{1} + w_6^{-1}  +\underline{w_5^{-1} w_6^{-1} w_8^{-1}}+ \underline{w_1 w_2 w_3 w_4}+ w_5^{-1} w_6^{-1} \right)z_1 z_2 \  ,
\]
\medskip
where we have underlined the terms which survive the Higgsing. The ones that do not survive can be immediately determined, they are the ones containing edges corresponding to fields with non-zero vevs.

In terms of \eqref{ineq}, this means the weights $w_j \rightarrow \Lambda^{k_j}$ must be such that $k_3 = 0 = -k_5-k_6-k_8 = k_1+k_2+k_3+k_4$ coming from the underlined terms and that they must all be strictly greater than any of $\{ k_1+k_2+k_3 \ , k_3+k_4 \ , -k_6 \ , -k_5 - k_6\}$ coming from the non-underlined terms.
This thus constitutes one of the inequalities.
We do this for each of the 8 points in the toric diagram and combine all these relations, supplementing by the inequality that $\sum\limits_{j=1}^8 k_j = 0$, and solve the resulting system over the integers. We will find precisely the first row of the solution table in \eqref{w-2F0}.
In the table below, we will also include, for reference, the final leading order weight for the surviving terms. For the $(1,1)$ term above, this is just $0$, whence the entry $\Lambda^0 = 1$ in the third column. These results are in full agreement with those derived using \eref{w-2F0} and \eref{Lambda_weight_w}. In the examples that follow, we have independently determined the $\Lambda$-scalings using both methods and confirmed their agreement.

\bigskip

\beq
\begin{array}{|c|c|c|}
\hline
\ \ \ (n_1,n_2) \ \ \ & \mbox{Loops} & \\
\hline
\hline
(0,0) & \underline{1} & 1 \\
\hline
(1,0) & 1+ \underline{w_1 w_2 w_5 w_6} & \Lambda^2\\
\hline
(2,0) & \underline{w_1 w_2 w_5 w_6} & \Lambda^2  \\
\hline
(1,1) & w_1 w_3 w_4+ w_3 w_4+\underline{w_3}+ \underline{1} + w_6^{-1} & 1 \\
 &  +\underline{w_5^{-1} w_6^{-1} w_8^{-1}}+ \underline{w_1 w_2 w_3 w_4}+ w_5^{-1} w_6^{-1} & \\
\hline
(2,1) & \ \ w_1 w_3+\underline{w_1 w_8^{-1}}+ \underline{w_8^{-1}}+ w_1+ w_6^{-1} w_8^{-1} \ \ & \Lambda \\
 & + w_6^{-1} w_7^{-1} w_8^{-1}+ \underline{w_1 w_2 w_3 w_5}+ \underline{w_1 w_2 w_3} & \\
\hline
(1,2) & \underline{w_3 w_4 w_5^{-1} w_6^{-1}} & \ \ \Lambda^{-2} \ \ \\
\hline
(2,2) & w_1 w_3 w_4 w_6^{-1}+ \underline{w_3 w_5^{-1} w_6^{-1} w_8^{-1}} & 1 \\
\hline
(3,2) & \underline{w_1 w_3 w_6^{-1} w_8^{-1}} & 1 \\
\hline 
\end{array}
\label{coefficients_2_F0_Higgsing_1}
\eeq

\vspace{1.5cm}

\noindent{\bf Higgsing 2:}
We can now perform a similar analysis for the second Higgsing and obtain:

\beq
\begin{array}{|c|c|c|}
\hline
\ \ \ (n_1,n_2) \ \ \ & \mbox{Loops}  & \\
\hline
\hline
(0,0) &  \underline{1} & 1 \\
\hline
(1,0) & 1+ \underline{w_1 w_2 w_5 w_6} & \Lambda^2 \\
\hline
(2,0) & \underline{w_1 w_2 w_5 w_6} & \Lambda^2 \\
\hline
(1,1) & \underline{w_1 w_3 w_4} + w_3 w_4+w_3+ \underline{1} + \underline{w_6^{-1}} & 1 \\
 & + w_5^{-1} w_6^{-1} w_8^{-1}+ \underline{w_1 w_2 w_3 w_4}+ w_5^{-1} w_6^{-1} & \\
\hline
(2,1) & \ \ w_1 w_3+ \underline{w_1 w_8^{-1}} + w_8^{-1}+ \underline{w_1}+ w_6^{-1} w_8^{-1} \ \ & \Lambda \\
 & + \underline{w_6^{-1} w_7^{-1} w_8^{-1}} + \underline{w_1 w_2 w_3 w_5}+ w_1 w_2 w_3 & \\
\hline
(1,2) & \underline{w_3 w_4 w_5^{-1} w_6^{-1}} & \ \ \Lambda^{-2} \ \ \\
\hline
(2,2) & \underline{w_1 w_3 w_4 w_6^{-1}}+w_3 w_5^{-1} w_6^{-1} w_8^{-1} & 1 \\
\hline
(3,2) & \underline{w_1 w_3 w_6^{-1} w_8^{-1}} & 1 \\
\hline 
\end{array}
\eeq

\vspace{2cm}

\noindent{\bf Higgsing 3:}
So too we can now study the third Higgsing, confirming our results:

\smallskip

\beq
\begin{array}{|c|c|c|}
\hline
\ \ \ (n_1,n_2) & \mbox{Loops} & \\
\hline
\hline
(0,0) & \underline{1} & 1 \\
\hline
(1,0) & \underline{1}+w_1 w_2 w_5 w_6 & 1 \\
\hline
(2,0) & \underline{w_1 w_2 w_5 w_6} & \ \ \Lambda^{-2} \ \ \\
\hline
(1,1) & \underline{w_1 w_3 w_4}+\underline{w_3 w_4}+w_3+1 + \underline{w_6^{-1}} & \Lambda \\
 & + w_5^{-1} w_6^{-1} w_8^{-1}+ w_1 w_2 w_3 w_4+ \underline{w_5^{-1} w_6^{-1}} & \\
\hline
(2,1) & \ \ \underline{w_1 w_3}+ w_1 w_8^{-1}+ w_8^{-1}+\underline{w_1}+ \underline{w_6^{-1} w_8^{-1}} \ \ & 1 \\
 & + \underline{w_6^{-1} w_7^{-1} w_8^{-1}}+ w_1 w_2 w_3 w_5+ w_1 w_2 w_3 & \\
\hline
(1,2) & \underline{w_3 w_4 w_5^{-1} w_6^{-1}} & \Lambda^2 \\
\hline
(2,2) & \underline{w_1 w_3 w_4 w_6^{-1}}+w_3 w_5^{-1} w_6^{-1} w_8^{-1} & \Lambda^2\\
\hline
(3,2) & \underline{w_1 w_3 w_6^{-1} w_8^{-1}} & 1 \\
\hline 
\end{array}
\eeq

\vspace{.8cm}

\noindent{\bf Higgsing 4:}
Finally, we complete the story with the last Higgsing:

\smallskip

\beq
\begin{array}{|c|c|c|}
\hline
\ \ \ (n_1,n_2) \ \ \ & \mbox{Loops} & \\
\hline
\hline
(0,0) & \underline{1} & 1 \\
\hline
(1,0) & \underline{1} + w_1 w_2 w_5 w_6 & 1 \\
\hline
(2,0) & \underline{w_1 w_2 w_5 w_6} & \ \ \Lambda^{-2} \ \ \\
\hline
(1,1) & w_1 w_3 w_4+  \underline{w_3 w_4}+ \underline{w_3}+1 + w_6^{-1} & \Lambda \\
 &  +  \underline{w_5^{-1} w_6^{-1} w_8^{-1}}+ w_1 w_2 w_3 w_4+  \underline{w_5^{-1} w_6^{-1}} & \\
\hline
(2,1) &  \ \ \underline{w_1 w_3}+ w_1 w_8^{-1}+ \underline{w_8^{-1}}+ w_1+\underline{ w_6^{-1} w_8^{-1}} \ \ & 1 \\
 & + w_6^{-1} w_7^{-1} w_8^{-1}+ w_1 w_2 w_3 w_5+\underline{w_1 w_2 w_3} & \\
\hline
(1,2) & \underline{w_3 w_4 w_5^{-1} w_6^{-1}} & \Lambda^2 \\
\hline
(2,2) & w_1 w_3 w_4 w_6^{-1}+\underline{w_3 w_5^{-1} w_6^{-1} w_8^{-1}} & \Lambda^2 \\
\hline
(3,2) & \underline{w_1 w_3 w_6^{-1} w_8^{-1}} & 1 \\
\hline 
\end{array}
\eeq

\bigskip

\subsubsection*{Am{\oe}ba Projections}

As reviewed in Section \ref{section_splitting_spectral_curves}, am{\oe}ba plots provide a simple way of visualizing the spectral curves. Thus, we can draw such projections of the spectral curve of the parent for some large value of $\Lambda$, as an additional check that the scalings we obtained indeed give rise to elongations leading to the desired splitting of the spectral curve. 

\begin{figure}[h]
\begin{center}
\includegraphics[width=9.5cm]{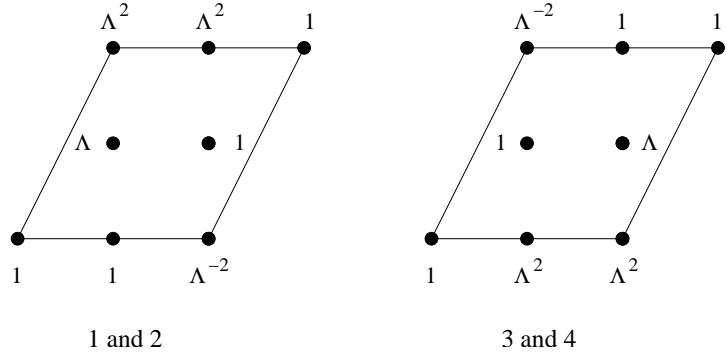}
\caption{The corresponding weights of the coefficients in the Newton polynomial for Higgsings 1 and 2 are connected to those of Higgsings 3 and 4 by a 180$^{\circ}$ rotation of the corresponding toric diagram.}
\label{toric_2_F0s_scaling}
\end{center}
\end{figure}

Let the $\Lambda$-weights for each node in the toric diagram be given by the third column in the tables above. The leading behavior in $\Lambda$ coincides for Higgsings 1 and 2 and for Higgsings 3 and 4.
Furthermore, the two pairs are related to each other by a $180^\circ$ rotation, as shown in 
\fref{toric_2_F0s_scaling}.

\begin{figure}[h]
 \centering
 \begin{tabular}[c]{ccc}
 \epsfig{file=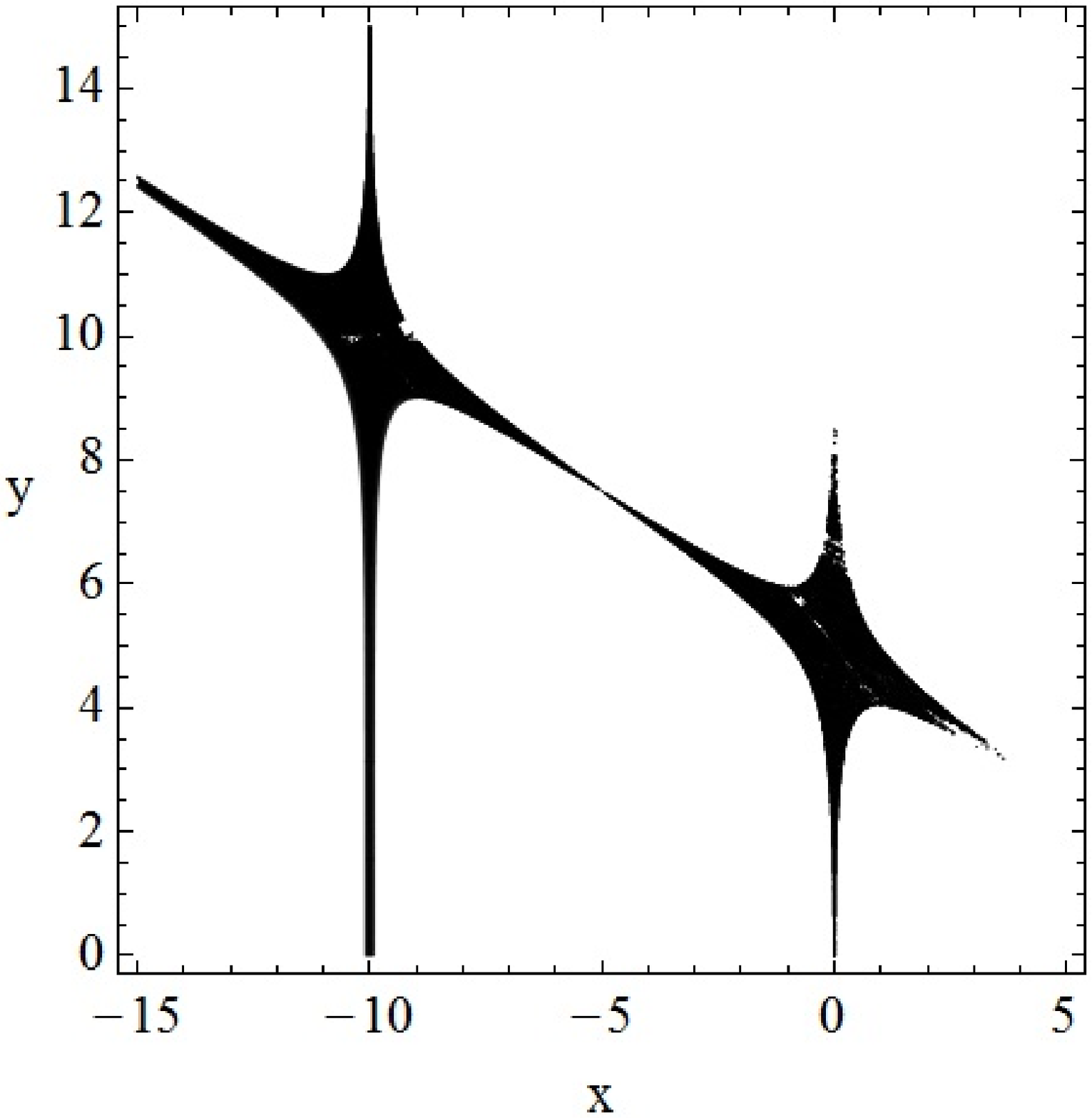,width=0.35\linewidth,clip=} & \ \ \ \ \ \ &
\epsfig{file=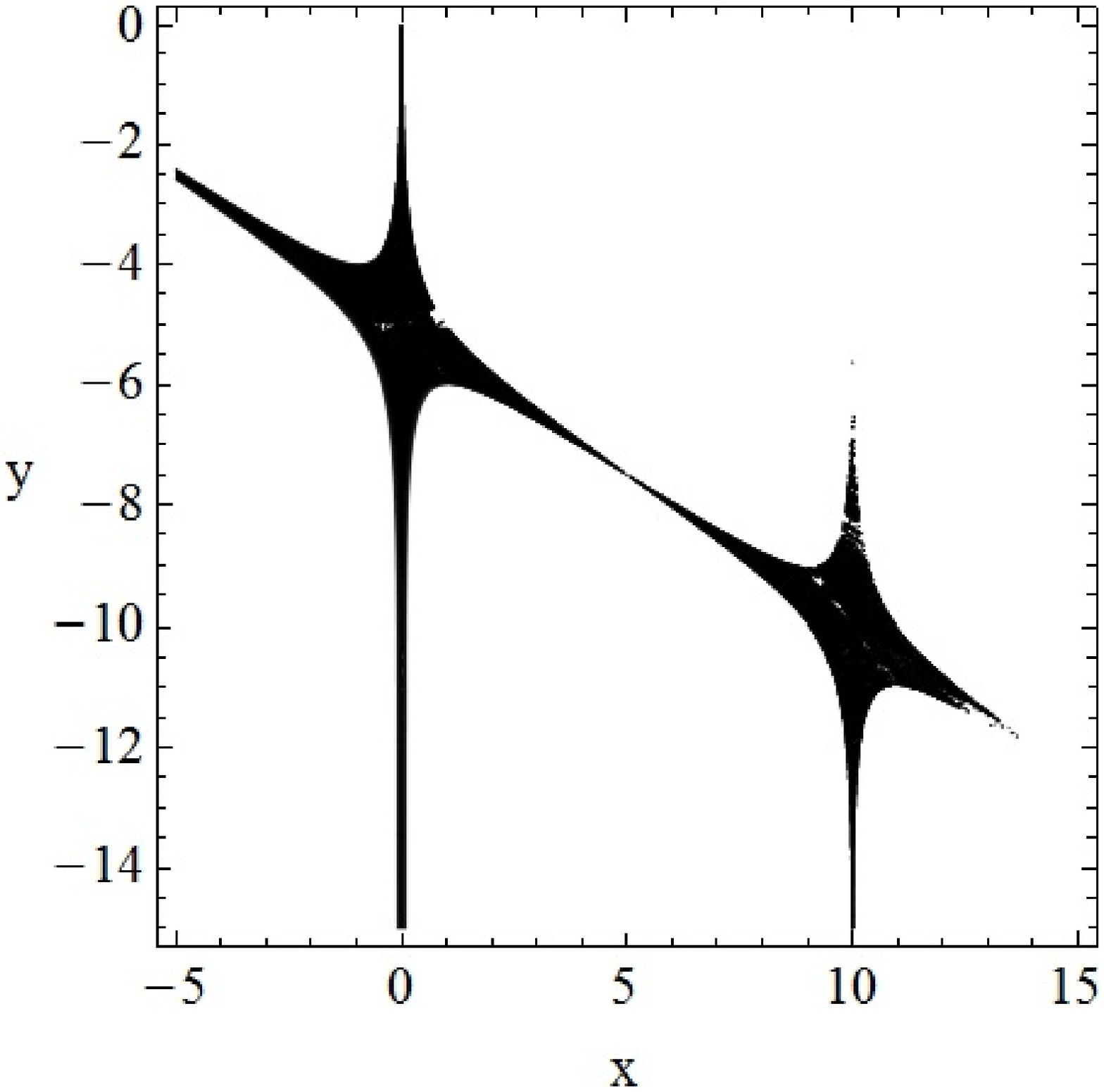,width=0.35\linewidth,clip=} \\
\mbox{Higgsings 1 and 2} & & \mbox{Higgsings 3 and 4} 
 \end{tabular}
\caption{Am{\oe}ba plots for the 4 possible Higgsings with $\Lambda$ set to the numerical value of $e^5$. The patchy appearance of these and subsequent am{oe}ba plots, with some missing points in their interior, is due to the fact that we determine them numerically.}
\label{amoebas_2F0} 
\end{figure}

We see that, in perfect agreement, the am{\oe}ba projections exhibit the corresponding behavior.
\fref{amoebas_2F0} shows the am{\oe}bas for the four Higgsings for large $\Lambda$.
Indeed, the am{\oe}bas for the four models coincide, up to a trivial shift on the $(x,y)$ plane. This results from the simple relation between their scalings as given by \fref{toric_2_F0s_scaling}. Furthermore, we see that the thin spine in the center controls precisely the spliting of the double $F_0$ into her two daughter $F_0$ theories. The holes in the spectral curve associated to internal points in the toric diagram have zero size in \fref{amoebas_2F0}. This is due to the particular choice of coefficients in the characteristic polynomial. These coefficients can be varied at will without modifying their $\Lambda$-scaling and hence preserving the splitting.

\bigskip

\subsection{$Y^{4,0}$}

\begin{figure}[h]
\begin{center}
\includegraphics[width=4cm]{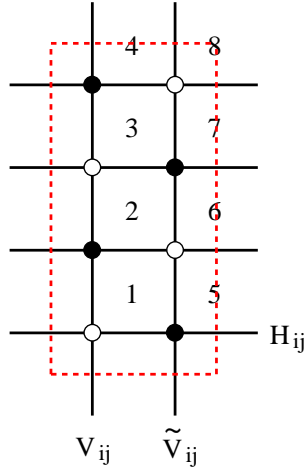}
\caption{Brane tiling for $Y^{4,0}$.}
\label{dimer_Y40}
\end{center}
\end{figure}

\begin{figure}[h]
 \centering
 \begin{tabular}[c]{ccc}
 \epsfig{file=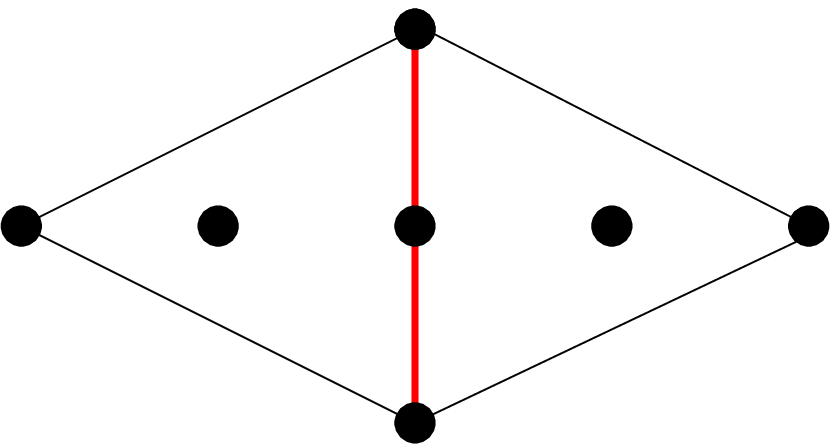,width=0.25\linewidth,clip=} & \ \ \ \ \ \ \ \ &
\epsfig{file=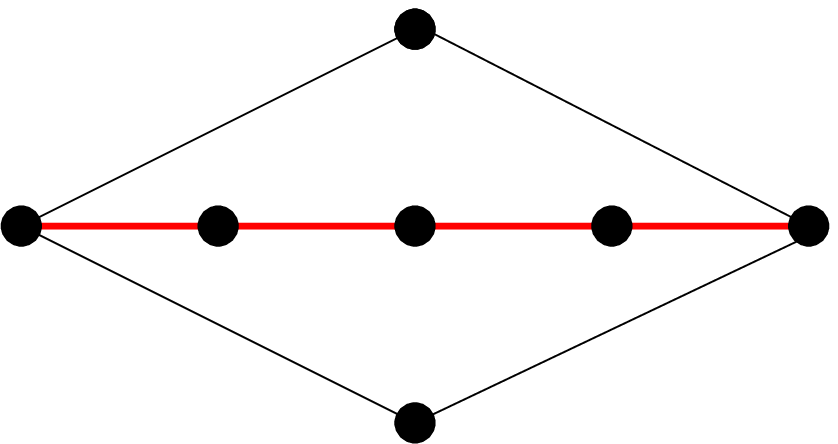,width=0.25\linewidth,clip=} \\
\mbox{Higgsing 1} & & \mbox{Higgsing 2} 
 \end{tabular}
\caption{Toric diagram for $Y^{4,0}$, showing the two splittings that we will investigate.
The splitting is indicated by the red line.}
\label{toric_Y40_Higgsings} 
\end{figure}

Fortified by the consistency our story for double $F_0$, let us move onto another non-trivial example.
We now consider $Y^{4,0}$, whose corresponding integrable system has been worked out in \cite{Eager:2011dp}. 
We remind the reader of the dimer model in \fref{dimer_Y40}.
This is a theory with 8 gauge group factors, quartic superpotential terms and 16 fields, which we suggestively label as $V_{ij}$, $\tilde{V}_{ij}$ and $H_{ij}$ in accordance with their vertical and horizontal orientation.
The toric diagram is shown in \fref{toric_Y40_Higgsings}, with 7 nodes.

We now study the two splittings depicted in \fref{toric_Y40_Higgsings}.
We will refer to them as Higgsings 1 and 2. Specifically, the acquisition of vevs is as follows:

\beq
\begin{array}{ccccccc}
\mbox{{\bf \underline{Higgsing 1}:}} & & H_{32}^{(1)}, H_{65}^{(1)}, H_{14}^{(1)}, H_{87}^{(1)} & \ \ \ \ & \mbox{{\bf \underline{Higgsing 2}:}} & & V_{51}^{(1)}, V_{73}^{(1)}, \tilde{V}_{26}^{(1)}, \tilde{V}_{48}^{(1)} \\
& & H^{(2)}_{34} , H^{(2)}_{67} , H^{(2)}_{12} , H^{(2)}_{85} & & & & \tilde{V}_{51}^{(2)}, \tilde{V}_{73}^{(2)}, V_{26}^{(2)}, V_{48}^{(2)}
\end{array}
\label{vev_Higgsings_Y40}
\eeq

We remark that the black and white nodes are exchanged with respect to \cite{Eager:2011dp}. This is of course only a matter of convention.
As in the previous example, the choice of vevs leading to each splitting is not unique. Having already illustrated this possibility with the splitting of double $F_0$ to two $F_0$'s, we focus on the vevs given in \eqref{vev_Higgsings_Y40} for convenience.

Once again, we determine the $\Lambda$-scalings associated to both decompositions using the methods of Section \ref{section_continuous_parameter}, i.e. both the rule in \eqref{Lambda_weight_w} and the algorithm in Section \ref{s:algo}, which yield identical results. We obtain:
\beq
\begin{array}{c|cccccccc}
\ \ \mbox{Higgsing} \ \ & \ \ w_1 \ \ & \ \ w_2 \ \ & \ \ w_3 \ \ & \ \ w_4 \ \ & \ \ w_5 \ \ & \ \ w_6 \ \ & \ \ w_7 \ \ & \ \ w_8 \ \ \\ \hline \hline
\mbox{b} & \Lambda^{-1} & \Lambda & \Lambda^{-1} & \Lambda & \Lambda & \Lambda^{-1} & \Lambda & \Lambda^{-1} \\ \hline
\mbox{c} & \Lambda & \Lambda^{-1} & \Lambda & \Lambda^{-1} & \Lambda^{-1} & \Lambda & \Lambda^{-1} & \Lambda 
\end{array}
\eeq

\paragraph{Higgsing 1:} We follow the notation in the previous examples, write the coefficients in terms of the $w$-monomials and underline the terms which survive.
We tabulate this for each node, and in the third column write the overall leading behavior in $\Lambda$ for the terms which survive.

\medskip

\beq
\begin{array}{|c|c|c|}
\hline
\ \ (n_1,n_2) \ \ & \mbox{Loops} & \\
\hline
\hline
(0,0) & \underline{1} & 1 \\
\hline
(-1,0) & \underline{w_4} + w_4 w_8+\underline{w_4 w_7 w_8}+ w_3 w_4 w_7 w_8 & \Lambda \\ &  +\underline{w_1^{-1} w_5^{-1} w_6^{-1}}+ w_1^{-1} w_5^{-1}+\underline{w_1^{-1}} +1 & \\
\hline
(-2,0) & w_1^{-1} w_5^{-1} w_4+ w_4 w_8+w_1^{-1} w_4 w_8+ w_1^{-1} w_5^{-1} w_4 w_8 & \Lambda^2 \\
& + w_1^{-1} w_5^{-1} w_6^{-1} w_4 w_8+ w_4 w_7 w_8+\underline{w_1^{-1} w_4 w_7 w_8}+ w_1^{-1} w_5^{-1} w_4 w_7 w_8 & \\
&+w_3 w_4 w_7 w_8+w_1^{-1} w_3 w_4 w_7 w_8+w_1^{-1} w_5^{-1} w_3 w_4 w_7 w_8+w_1^{-1} w_5^{-1} w_6^{-1} & \\ & + w_1^{-1} w_5^{-1}+w_3 w_4^2 w_7 w_8+ \underline{w_4 w_1^{-1} w_5^{-1} w_6^{-1}}+w_3 w_4^2 w_7 w_8^2 & \\
\hline
(-3,0) & \ w_1^{-1} w_5^{-1} w_4 w_8+\underline{ w_1^{-1} w_5^{-1} w_6^{-1} w_4 w_8}+ \underline{w_1^{-1} w_5^{-1} w_4 w_7 w_8} \ & \Lambda \\ & + w_1^{-1} w_5^{-1} w_3 w_4 w_7 w_8+ \underline{w_1^{-1} w_5^{-1} w_3 w_4^2 w_7 w_8}+ w_3 w_4^2 w_7 w_8^2& \\ & + \underline{w_1^{-1} w_3 w_4^2 w_7 w_8^2}+ w_1^{-1} w_5^{-1} w_3 w_4^2 w_7 w_8^2 & \\
\hline
(-4,0) & \underline{w_1^{-1} w_5^{-1} w_3 w_4^2 w_7 w_8^2} & 1 \\
\hline
(-2,1) & \underline{w_1^{-1} w_4 w_7 w_8} & \Lambda^2 \\
\hline
(-2,-1) & \underline{w_2 w_3 w_4^2 w_7 w_8} & \ \Lambda^2 \ \\
\hline
\end{array}
\label{scalings_Y40_Higgsing_1}
\eeq

\bigskip

\paragraph{Higgsing 2:}
Likewise, we tabulate the result for Higgsing 2 and obtain:

\medskip

\beq
\begin{array}{|c|c|c|}
\hline
\ \ (n_1,n_2) \ \ & \mbox{Loops} & \\
\hline
\hline
(0,0) & \underline{1} & 1 \\
\hline
(-1,0) & w_4 + \underline{w_4 w_8}+w_4 w_7 w_8+ \underline{w_3 w_4 w_7 w_8} & 1 \\ &  +w_1^{-1} w_5^{-1} w_6^{-1} +\underline{w_1^{-1} w_5^{-1}}+w_1^{-1} +\underline{1} & \\
\hline
(-2,0) & w_1^{-1} w_5^{-1} w_4+ \underline{w_4 w_8}+w_1^{-1} w_4 w_8+ \underline{w_1^{-1} w_5^{-1} w_4 w_8} & 1 \\
& + w_1^{-1} w_5^{-1} w_6^{-1} w_4 w_8+ w_4 w_7 w_8+ w_1^{-1} w_4 w_7 w_8+ w_1^{-1} w_5^{-1} w_4 w_7 w_8 & \\
&+ \underline{w_3 w_4 w_7 w_8}+w_1^{-1} w_3 w_4 w_7 w_8+\underline{w_1^{-1} w_5^{-1} w_3 w_4 w_7 w_8}+w_1^{-1} w_5^{-1} w_6^{-1} & \\ & + \underline{w_1^{-1} w_5^{-1}}+w_3 w_4^2 w_7 w_8+w_4 w_1^{-1} w_5^{-1} w_6^{-1}+\underline{w_3 w_4^2 w_7 w_8^2 } & \\
\hline
(-3,0) & \underline{w_1^{-1} w_5^{-1} w_4 w_8}+ w_1^{-1} w_5^{-1} w_6^{-1} w_4 w_8+ w_1^{-1} w_5^{-1} w_4 w_7 w_8 & 1 \\ & + \underline{w_1^{-1} w_5^{-1} w_3 w_4 w_7 w_8}+ w_1^{-1} w_5^{-1} w_3 w_4^2 w_7 w_8+ \underline{w_3 w_4^2 w_7 w_8^2} & \\ & +w_1^{-1} w_3 w_4^2 w_7 w_8^2+ \underline{w_1^{-1} w_5^{-1} w_3 w_4^2 w_7 w_8^2} & \\
\hline
(-4,0) & \underline{w_1^{-1} w_5^{-1} w_3 w_4^2 w_7 w_8^2} & 1 \\
\hline
(-2,1) & \underline{w_1^{-1} w_4 w_7 w_8} & \ \Lambda^{-2} \ \\
\hline
(-2,-1) & \underline{w_2 w_3 w_4^2 w_7 w_8} & \Lambda^{-2} \\
\hline
\end{array}
\label{scalings_Y40_Higgsing_2}
\eeq

\bigskip

\begin{figure}[h]
 \centering
 \begin{tabular}[c]{ccc}
 \epsfig{file=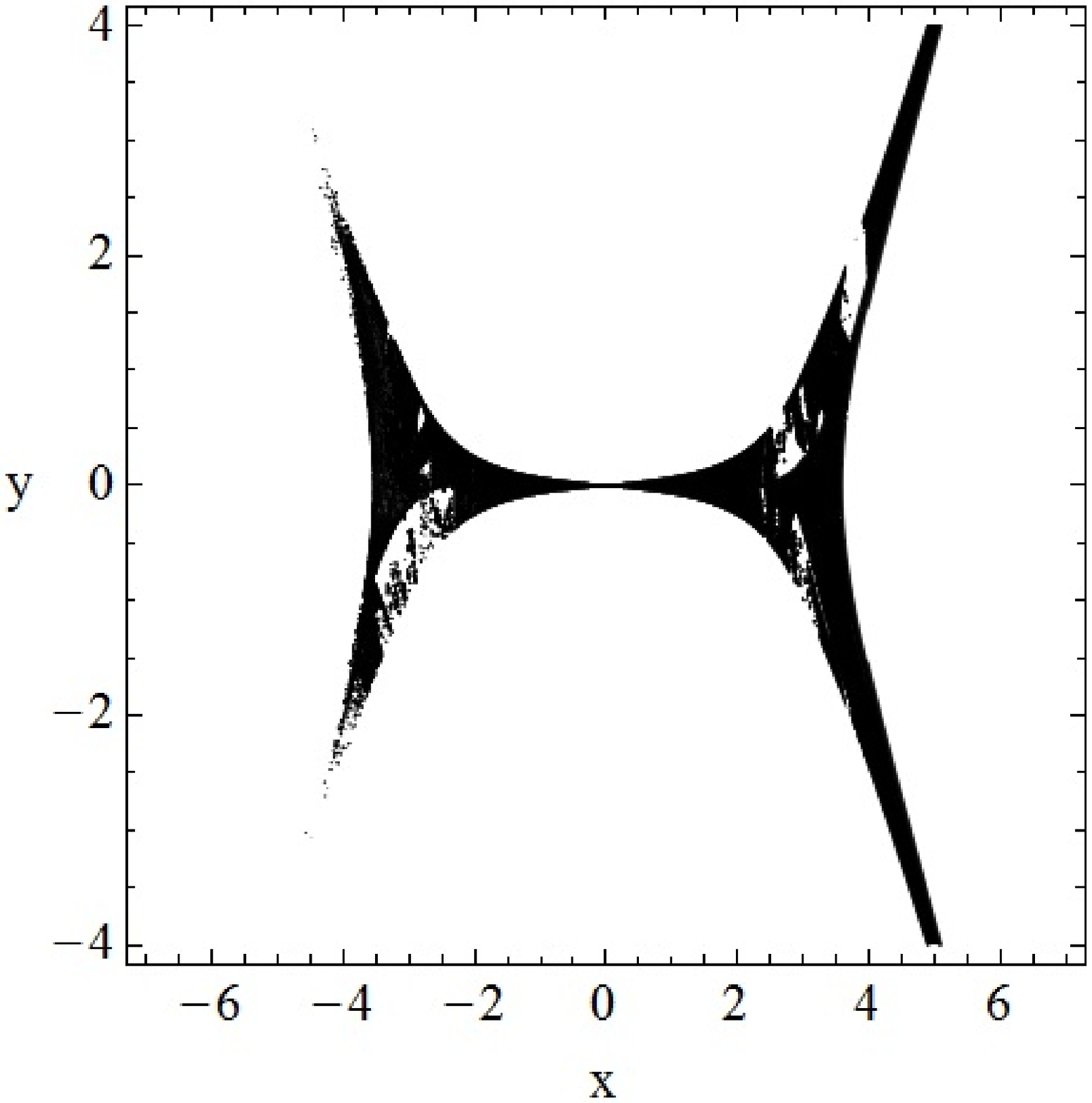,width=0.35\linewidth,clip=} & \ \ \ \ \ \ &
\epsfig{file=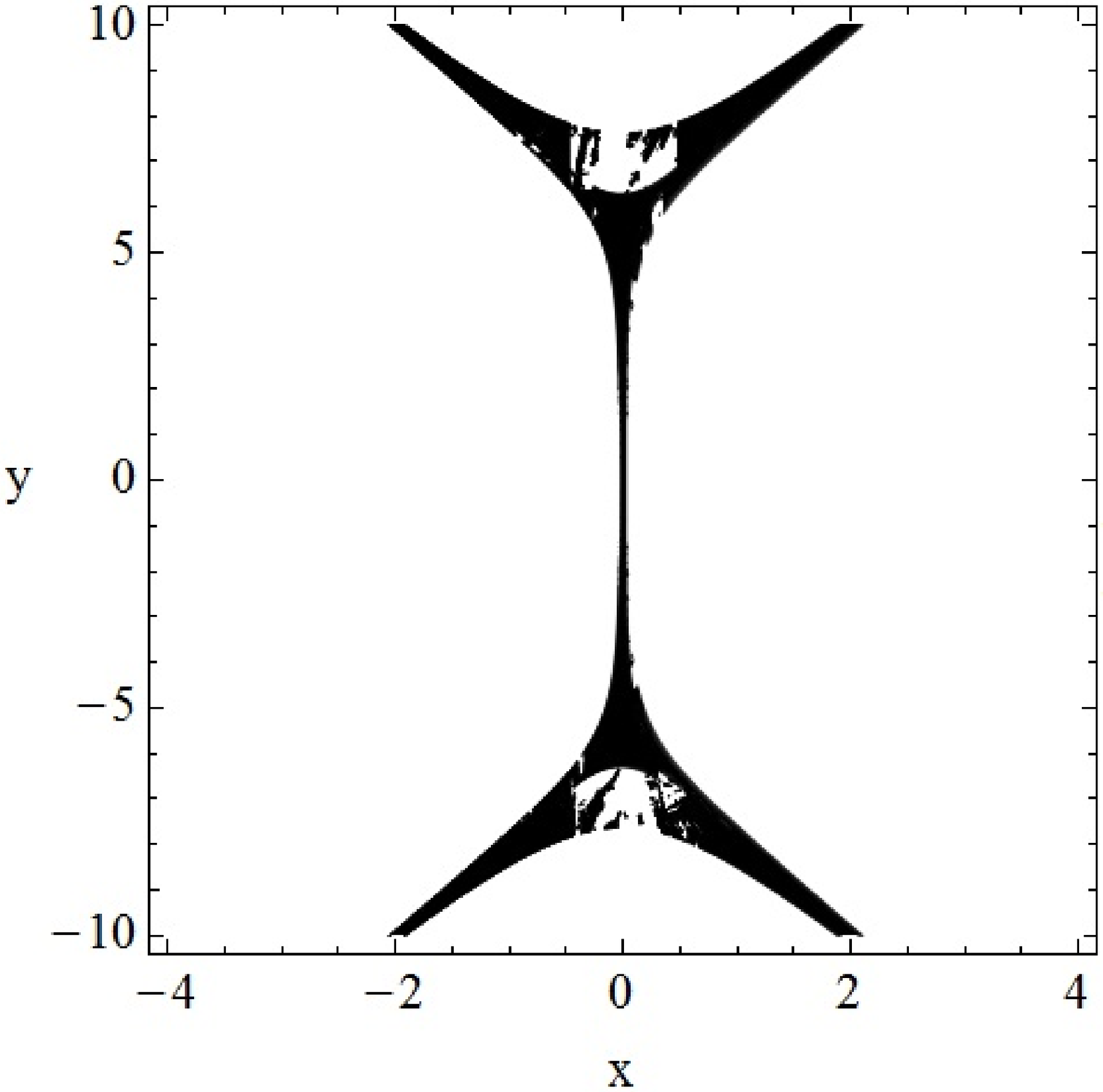,width=0.35\linewidth,clip=} \\
\mbox{Higgsing 1} & & \mbox{Higgsing 2} 
 \end{tabular}
\caption{Am{\oe}ba plots for Higgsings 1 and 2 of $Y^{4,0}$ at $\Lambda=e^3$.}
\label{amoebas_Y40} 
\end{figure}

\fref{amoebas_Y40} shows the am{\oe}bas corresponding to the scalings in \eref{scalings_Y40_Higgsing_1} and \eref{scalings_Y40_Higgsing_2}, confirming they produced the desired splitting.

\section{Combining Multiple Components}

\label{section_large-N}

In the previous sections, we have explained how to split integrable systems. Reversing the logic, we also understand how to glue them. We have identified a continuous parameter $\Lambda$ that controls the distance between components of the spectral curve.  This parameter manifests itself in the associated quivers as non-zero vevs and suppresses certain contributions in the integrable system. We can now proceed towards our goal of understanding the continuous limit of these systems.

\medskip

\subsection{Combinatorics of a Large Number of Components}

Let us focus on the case in which we combine an infinite number of identical components $\Sigma_0$ along a single direction, effectively generating a new continuous dimension. The am{\oe}ba projection suggests a natural way to approach the continuum: we consider all components equally separated in the am{\oe}ba and then send the number of components $N$ contained in a finite interval of length $L$ to infinity. More concretely, defining $\Lambda\equiv e^\alpha$, we consider the limit
\beq
N\to \infty, \ \ \ \ \ \ \ \alpha\to 0, \ \ \ \ \ \ \ L=N \alpha \mbox{ fixed}.
\eeq
This will give us a $(1+1)$-dimensional integrable system from the continuous limit of an infinite number of $(0+1)$-dimensional ones. In principle, it seems possible to do the same in both the $x$ and $y$ directions, generating a $(2+1)$-dimensional integrable field theory in the process.

Before studying this limit, let us investigate how the number of contributions to Hamiltonians behaves for large $N$. Following the dictionary in Section \ref{section_dimer_cluster_IS}, this number corresponds to the multiplicity of perfect matchings associated to internal points in the toric diagram.\footnote{Casimirs are given by ratios of external points in the toric diagram. Hence, the same ideas apply independently to their numerator and denominator.}

Gluing $N$ copies of an integrable system corresponds to considering a certain $\mathbb{Z}_N$ orbifold of the basic theory. In dimer model language, this corresponds to enlarging the unit cell by a factor $N$. While the number of points in the toric diagram grows linearly with $N$, their multiplicity grows much faster. Let us illustrate this growth in some explicit examples.

\medskip

\subsubsection{$Y^{N,0}$}

\begin{figure}[h]
\begin{center}
\includegraphics[width=6.5cm]{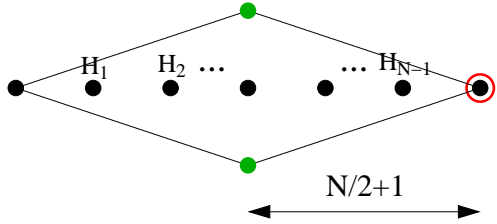}
\caption{Toric diagram for $Y^{N,0}$ for even $N$. The red circle indicates the reference perfect matching and the green dots correspond to cycles with windings $(-N/2 - 1, 1)$ and $(-N/2 - 1,-1)$, which are fixed by the Casimirs.}
\label{toric_general_Yp0}
\end{center}
\end{figure}

The cone over $Y^{N,0}$ is the $\mathbb{Z}_N$ orbifold of the conifold with toric diagram given by \fref{toric_general_Yp0}.

The integrable system for this geometry was determined in \cite{Eager:2011dp} using the prescription in \cite{GK}, where it was identified with the $N$-site relativistic periodic Toda chain. Our goal in this section is to investigate its behavior for large $N$.  

\begin{figure}[h]
\begin{center}
\includegraphics[width=13cm]{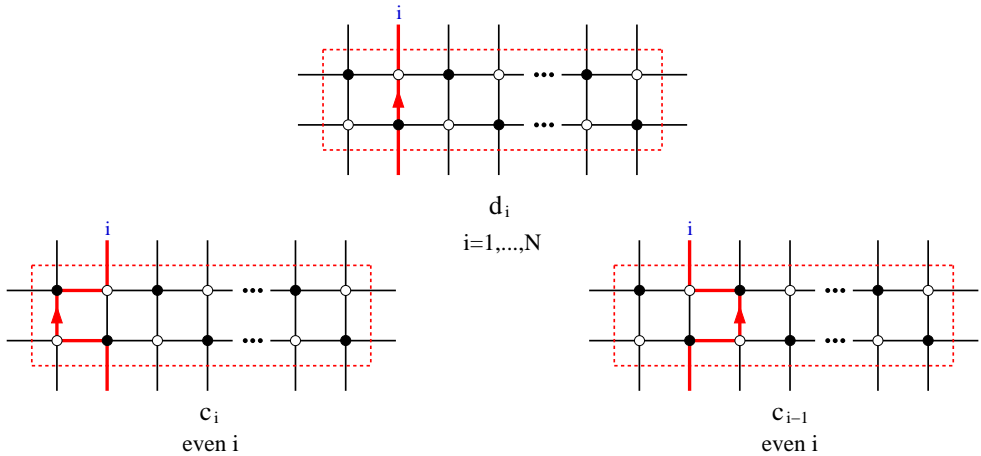}
\caption{A convenient set of cycles for $Y^{N,0}$ with even $N$. The cycles of type $c$ only exist for even $i$.}
\label{cycles_Yn0}
\end{center}
\end{figure}

For simplifity, let us focus on the case of even $N$. A similar analysis is possile for odd $N$. It turns out that the resulting integrable system is considerably simplified when considering the basis of cycles given in \fref{cycles_Yn0}, instead of using the standard $w$ and $z$-variables.\footnote{\fref{cycles_Yn0} shows $2N$ cycles. The two additional cycles that are necessary to form a basis, are fixed by the Casimirs and hence not important in our discussion. They correspond to the green dots in \fref{toric_general_Yp0}.} In terms of this basis, the Hamiltonians become

\beq
H_n=\sum \prod \underbrace{d_i \, c_j}_{\mbox{$n$ factors}}.
\label{Hn_Yp0}
\eeq
The problem of finding the Hamiltonians is thus reduced to the combinatorics of non-intersecting paths, which can be used to immediately determine the multiplicity of internal points in the toric diagram. A closed expression for this multiplicity was derived in \cite{OR} by using a Potts model like description for the dimers, and via a recursion relation that was obtained from a map to a 1-dimensional monomer-dimer system. The final result for the multiplicity of the $n^{th}$ internal point is

\beq
\sum_{i=0}^n {N \over N-i} \binom{n}{i} \binom{N-i}{n} \, ,
\eeq
which applies for both even and odd $N$.

\begin{figure}[h]
\begin{center}
\includegraphics[width=8cm]{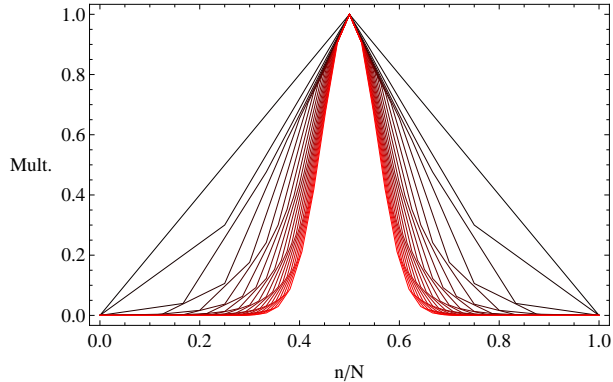}
\caption{Normalized multiplicity of perfect matchings for the internal points of the $Y^{N,0}$ theory. We have also normalized the length of the toric diagram to $1$. We show results for $N=2a$, $a=1,\ldots,20$ (black to red).}
\label{multiplicity_YN0}
\end{center}
\end{figure}

It is interesting to visualize how these multiplicities are distributed over the toric diagram and how the distribution approaches some limit shape after appropriate normalization. This is shown in \fref{multiplicity_YN0}.

\subsubsection{Multiple $F_0$}

\begin{figure}[h]
\begin{center}
\includegraphics[width=6.5cm]{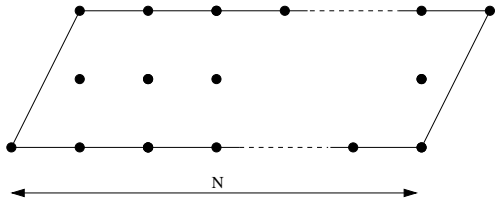}
\caption{Toric diagram for the $N$ $F_0$ model.}
\label{toric_N_F0}
\end{center}
\end{figure}

Similarly, we can investigate the generalization of the model considered in Section \ref{section_double_F0} to $N$ copies of $F_0$. \fref{toric_N_F0} shows its toric diagram.

In this case, it is also possible to find closed formulas for the multiplicities of all points in the toric diagram. They are

\beq
\mbox{Boundary points:} \ \binom{N}{m} 
\ \ \ \ \
\mbox{Internal points:} \ 2 \binom{2N}{2n-1}
\eeq
where $N$ is the number of $F_0$'s that have been glued together and $m \in \{0,N\}$ and $n \in \{1,N\}$ index points on the boundary and in the interior of the toric diagram, respectively.

\begin{figure}[h]
\begin{center}
\includegraphics[width=8cm]{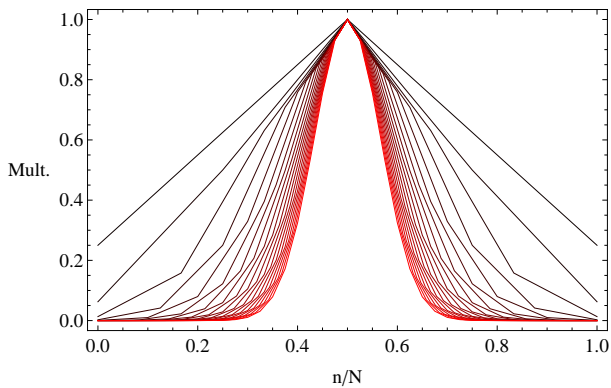}
\caption{Normalized multiplicity of perfect matchings for the internal points of the multiple $F_0$ theory. We have also normalized the length of the toric diagram to $1$. We show results for $N=2a+1$, $a=0,\ldots,20$ (black to red).}
\label{multiplicity_N_F0}
\end{center}
\end{figure}

These two examples illustrate a general behavior of large-$N$ models, an explosive growth in the number of perfect matchings associated to a given point in the toric diagram, which translated into a huge number of contributions for each conserved charge. It thus becomes clear that a continuous reformulation of cluster integrable systems is desirable in order to deal with their large-$N$ limit. In the next section we take the first steps towards such a reformulation.

\medskip

\subsection{A Toy Model for the Continuous Limit}

The two examples considered in the previous subsection share some common characteristics. In both of them, the $n^{th}$ Hamiltonian corresponds to the sum over all possible positions on the brane tiling of $n$ paths, subject to the constraint of not overlapping over edges. Furthermore, these paths are of a very specific type: they cross the tiling along the short direction of the unit cell and are {\it almost straight}. By this we mean that these paths are almost localized along the long direction of the unit cell. \fref{cycles_Yn0} shows the explicit form of these paths for $Y^{N,0}$ and \eref{Hn_Yp0} gives the corresponding Hamiltonians. It is natural to assume that this structure is generic when gluing $N$ copies of a cluster integrable system with a genus-$1$ spectral curve.\footnote{Attaching $N$ copies of a genus-$N$ spectral curve results in $g \, N$ Hamiltonians. We expect the resulting theories to obey a similar structure.}  In this section, we introduce a toy model with these properties, which we expect captures the main features of a continuous reformulation of cluster integrable systems.

Let us consider a system in the $x\in [0,L]$ interval and introduce a path $z$ winding vertically at $x=0$.

\begin{figure}[h]
\begin{center}
\includegraphics[width=9cm]{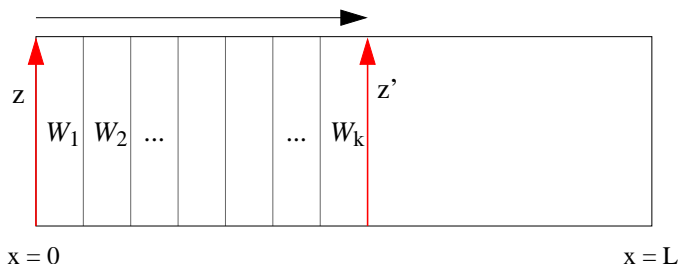}
\caption{A toy model for the continuous limit of cluster integrable systems. The path $z$ can be shifted by multiplying by all the $\mathcal{W}_i$'s contained in the strip between its initial and final positions.}
\label{toy_continuous}
\end{center}
\end{figure}

We can shift $z$ horizontally by multiplying it by all the $w_i$ variables contained in the strip between $x=0$ and its final position. 

\beq
z \to \prod_{i\in \mbox{strip}} \mathcal{W}_i \, z \, .
\eeq
Here $\mathcal{W}_i$ stands in general for a product of $w_i$'s contained in a slice of the tiling. The explicit form of $\mathcal{W}_i$ is controlled by the details of the specific brane tiling under consideration. Combining $n$ paths and summing over their positions $k_i$, $i=1,\ldots,n$, we obtain

\beq
H_n= \left( \prod_{i=1}^n \, \sum_{k_i < k_{i+1}} \prod_{j=0}^{k_i} \mathcal{W}_j \right) \, z^n \, ,
\eeq
which we have suggestively called $H_n$, since it has the expected structure for Hamiltonian operators.\footnote{Strictly speaking, Hamiltonians might also contain an $n$-independent power of the cycle orthogonal to $z$, as dictated by the position of the corresponding internal point in the toric diagram. This fact can be trivially incorporated in our expressions so, for simplicity, we omit it from our discussion.} We can readily generalize this expression in the continuous limit, obtaining

\beq
H_n= \left( \prod_{i=1}^n \int_{x_{i-1}}^{L} dx_i \, e^{\int_0^{x_i} dy \ln \mathcal{W}(y)} \right) \, z^n \, ,
\label{Hn_continuous}
\eeq
where $x_0\equiv 0$. While we have derived \eref{Hn_continuous} under rather basic assumptions that try to capture the most basic features observed in explicit examples, we expect it displays the main aspects of the actual continuous limit of cluster integrable systems. We leave a detailed investigation of this limit in explicit models for future work.

\bigskip

\section{Conclusions}

\label{section_conclusions}

We have taken the initial steps in extending the correspondence between dimer models and (0+1)-dimensional cluster integrable systems to continuous (1+1) and (2+1)-dimensional integrable theories. In order to understand the transition between discrete and continuous theories, it is necessary to have certain notion of distance between elementary constituents, or ``lattice spacing", such that the continuous theory emerges when it is sent to zero. We identified such a continuous parameter controlling the distance between daughters from the perspectives of both spectral curves and the resolution of Calabi-Yau singularities, equivalently the Higgsing of quivers. Furthermore, we introduced two procedures for determining the integrable system dependence on this parameter, whose effect is to suppress certain contributions to conserved charges, making them vanish in the infinite separation limit.

We then explored the integrable systems that are constructed by combining a large number of components, equivalently by gluing a large number of toric diagrams. More concretely, we studied, in explicit examples, the behavior of the number of contributions to individual Hamiltonians as the number of components grows. These contributions are in one-to-one correspondence with perfect matchings of the underlying dimer model. For this reason, their number diverges much more rapidly than the number of components, begging for an alternative continuous formulation of cluster integrable systems. We also observed that each Hamiltonian is given by the contributions of a number of simple paths summed over all their possible positions on the brane tiling. We used these insights to develop a toy model that we expect reproduces the basic features of the continuous limit of cluster integrable systems.

Interestingly, our investigation of the continuous merging of integrable system has also resulted in a novel understanding of (un)Higgsing in quiver theories and the associated desingularization of the corresponding Calabi-Yau spaces. Thus, we have added a new angle of attack to the classical subject of D-brane resolution of singularities. We have realized that when one refines the coefficients of 
the spectral curve into polynomials in loop variables, the Higgsing/resolution simply corresponds to establishing a consistent scale $\Lambda$ dictating which monomials should survive or suppressed. Therefore, we have effectively generated a new algorithm, outlined in Section \ref{s:algo}, for systematically studying all partial resolutions for a given toric diagram. It would be worthwhile to exploit this procedure for classifying all consistent daughter theories for a given parent.

What we have touched upon is, of course, only the beginning of a program. The natural question that arises now is how to extend our continuous toy model to theories that are actually constructible from dimers. For example, attempting to recover simple integrable field theories such as Toda theories would be an obvious next step. Given the simplifications afforded by dimer models, we expect it should be possible to construct increasingly more elaborate integrable field theories.

We envision many applications of dimer models to continuous theories, such as the study of integrablity preserving lower dimensional impurities or interfaces between different integrable field theories. Indeed, in light of the correspondence with dimer models, the often difficult condition of integrability simply amounts to checking whether the field theory results from the infinite limit of consistent toric diagrams, i.e. that they are given by convex lattice polygons.

In addition to the continuous limit, there are several exciting directions worth studying for the dimer model/integrable system correspondence. For example, the correspondence naturally associates (0+1)-dimensional relativistic integrable systems to the (3+1)-dimensional superconformal field theories dual to the dimer models. Recently, similar integrable systems have emerged in the study of (3+1)-dimensional ${\cal N}=2$ superconformal theories in the contexts of superconformal indices \cite{Gadde:2011uv} and the enumeration of vacua in the Omega background \cite{Nekrasov:2009rc,Nekrasov:2010ka}. It would be interesting to investigate whether the different ways in which integrable systems emerge are indeed related.

\bigskip

\section*{Acknowledgments}

The work of S. F. was supported by the US DOE under contract number DE-AC02-76SF00515 and by the STFC. S. F. and D. G. were also supported by the U.K. Science and Technology Facilities Council (STFC). Y. H. H. would like to thank the Science and Technology Facilities Council, UK, for an Advanced Fellowship and grant ST/J00037X/1, the Chinese Ministry of Education, for a Chang-Jiang Chair Professorship at NanKai University, the US National Science Foundation for grant CCF-1048082, as well as City University, London and Merton College, Oxford, for their enduring support.


\bigskip


\end{document}